\def\USEACHEMSO{0} %

\if\USEACHEMSO1
\documentclass[journal=jctcce,manuscript=article,%
email=false,%
]{achemso}
\else
\documentclass[aip,notitlepage,%
jcp,
tightenlines,
twocolumn,
reprint,%
floatfix]{revtex4-2} 
\fi

\usepackage[utf8]{inputenc}
\usepackage[T1]{fontenc} 
\usepackage[version=3]{mhchem}
\usepackage{hyperref}
\usepackage{units}

\usepackage{booktabs}
\usepackage{dcolumn}%
\usepackage{longtable}
\usepackage{geometry}
\usepackage{xspace}
\usepackage{verbatim}
\usepackage{amsmath}
\usepackage{amssymb}
\usepackage{bm}
\usepackage{subfigure}
\let\oldtheequation\theequation
\makeatletter
\def\tagform@#1{\maketag@@@{\ignorespaces#1\unskip\@@italiccorr}}
\renewcommand{\theequation}{(\oldtheequation)}
\makeatother 

\if\USEACHEMSO1
\fi

\usepackage{algorithm}
\usepackage{algpseudocode}
\makeatletter
\def\BState{\State\hskip-\ALG@thistlm}
\makeatother

\usepackage{morefloats}

\usepackage{listings}

\usepackage{xcolor,graphicx}

\newcommand{\matr}[1]{\ensuremath{\mathbf{#1}}}
\newcommand{\braket}[2]{\ensuremath{ \langle #1 | \, #2  \rangle }}

\newcommand{\ketbra}[2]{\ensuremath{  | {#1} \rangle \langle {#2} |}}

\newcommand{\ket}[1]{\ensuremath{  | {#1} \rangle}}

\newcommand{\matrixe}[3]{\ensuremath{ \langle{#1} | {#2} | {#3} \rangle }}

\newcommand{\lit}[1]{Ref.~\mbox{[\!\!\citenum{#1}]}\xspace}
\newcommand{\lits}[1]{Refs.~\mbox{[\!\!\citenum{#1}]}\xspace}

\definecolor{CBdblue}{RGB}{5,113,176}

\newcommand{\icm}{cm^{-1}}

\newcommand{\Nblock}{\ensuremath{N_\text{block}}}
\newcommand{\bdim}{\ensuremath{D}}
\newcommand{\bdimmax}{\ensuremath{D_\text{max}}}
\newcommand{\Nkrylov}{\ensuremath{N_\text{Krylov}}}
\newcommand{\tolkrylov}{\ensuremath{\epsilon_\text{Krylov}}}
\newcommand{\epsSVD}{\ensuremath{\epsilon_\text{SVD}}}

\if\USEACHEMSO0
\begin{document}
\fi

\title{Computing excited eigenstates using inexact Lanczos methods and tree tensor network states}
\author{Madhumita Rano}
\affiliation{Department of Chemistry and Biochemistry, University of California, Merced, CA 95343, USA}
\author{Henrik R.~Larsson}
\thanks{These authors contributed equally to this work.}
\email{planczos_25 [a t] larsson-research . $\delta$e}
\affiliation{Department of Chemistry and Biochemistry, University of California, Merced, CA 95343, USA}

\if\USEACHEMSO1
\begin{document}
\fi

\begin{abstract}
To understand the dynamics of quantum many-body systems, it is essential to study excited eigenstates. 
While tensor network states have become a standard tool for computing ground states in computational many-body physics, obtaining accurate excited eigenstates remains a significant challenge. 
In this work, we develop an approach that combines the inexact Lanczos method, which is designed for efficient computations of excited states, with tree tensor network states (TTNSs).
We demonstrate our approach by computing 
excited vibrational states for three challenging problems:
(1) 122
states in two different energy intervals of acetonitrile (12-dimensional),
(2) Fermi resonance states of the fluxional Zundel ion (15-dimensional),
and (3) selected excited states of the fluxional and very correlated Eigen ion (33-dimensional).
The proposed TTNS inexact Lanczos method is directly applicable to other quantum many-body systems.
\end{abstract}

\maketitle
\section{Introduction}
\label{sec:intro}
Vibrational spectra reveal deep insights into quantum effects of chemical bonding.\cite{Quantum2006huang,Vibrational2008wang,Fourth2012csaszar,CH52015wodraszka,Encoding2017devine,Stateresolved2022larsson,Coupling2022schroder,Smolyak2022chen,Chromium2022larsson,Quantum2023simko,Automated2024schroder,H2O2025simko}
Computing accurate spectra of strongly coupled or fluxional molecules is far from trivial, however,
particularly if detailed information of the spectroscopically active eigenstates is required.\cite{Variational2008bowman,Perspective2017carrington,Exact2023matyus,Benchmarking2025larsson}
One major challenge of computing eigenstate-resolved spectra is that many excited state computations are necessary in spectral regions that exhibit very high densities of states.\cite{vibronic2024larsson,Benchmarking2025larsson}
For standard basis set methods that lead to a linear eigenvalue problem, this challenge has been overcome in various ways.\cite{Efficient1994bramley,Extraction1995wall,Harmonic1998mandelshtam,New2000huang,Jacobi2000sleijpen,Accelerating2001poirier,HARMONIC2003mandelshtam,Solving2009gyorffy,New2017petrenko}
However, for more sophisticated methods that solve a nonlinear eigenvalue problem, computing excited states still poses a great challenge.\cite{Nonlinear2004mehrmann,Nonlinear2016vanbarel,Contour2023brennan} 
One example where computing excited states is difficult, particularly in a selected energy interval, is
tensor network state (TNS) methods, which have recently been used to compute eigenstates of complex systems to high accuracies.\cite{Iterative2014wang,Vibrational2017baiardi,Computing2019larsson,ExcitedState2021baiardi,Stateresolved2022larsson,Flexible2023glaser,vibronic2024larsson,Eigenstate2024hoppe,Benchmarking2025larsson} 
These methods include the multilayer multi-configurational time-dependent Hartree (ML-MCTDH) method,\cite{Multiconfiguration2000beck,Multilayer2015wang,Wavepacket2017manthe,Tensor2024larsson}
and methods based on the density matrix renormalization group (DMRG).\cite{Density1992white,Densitymatrix1993white,Tensor2024larsson}
Recently, we showed that DMRG methods 
that are based on a tree TNS (TTNS), which is the wavefunction ansatz of ML-MCTDH, 
can be used with great efficiency for accurately computing thousands of low-energy states for complex systems such as the fluxional, 15-dimensional Zundel ion and the strongly vibronically coupled \ce{NO3} radical. \cite{Computing2019larsson,Stateresolved2022larsson,vibronic2024larsson,Benchmarking2025larsson}
However, 
many TNS methods to target excited states directly are either costly or cannot target regions with a very high-density of states.\cite{Targeted2007dorando,Computing2019larsson,Optimization2019baiardi}

In general, for computing excited states, 
very efficient are 
spectral transformations of the Hamiltonian $\hat H$ that turn an interior excited state into an exterior state. 
Next to many possible transformations,\cite{Circumventing1994neuhauser,Shiftinvert1994meerbergen,General1996chen,HARMONIC2003mandelshtam,Variational2009matyus,Resolution2002barinovs} the shift-and-invert transformation, $(\hat H - \sigma \hat 1)^{-1}$, is one of the most common ones. For example, it is used in the Lanczos algorithm that we detail below.\cite{Spectral1980ericsson,Templates2000bai}
Lanczos and other algorithms based on this transformation
compute excited states near a target energy by
creating a compact basis, $\{\ket{\Psi_i}\}$, which is generated by solving linear equations of the type 
\begin{equation}
(\hat H - \sigma_i \hat I) \ket{\Psi_i} = \ket{\tilde \Psi}, \label{eq:SI_system}
\end{equation}
where  $\sigma_i$ is an energy close to the target, and the definition of $\ket{\tilde \Psi}$ depends on the particular algorithm. For example, $\ket{\tilde \Psi}$ can be $\ket{\Psi_{i-1}}$. 
In many cases, $\sigma_i$ does not depend on $i$, and the index is dropped.
Once the compact basis $\{\ket{\Psi_i}\}$ is generated,
the Hamiltonian $\hat H$ is projected onto the space spanned by the basis, and the matrix representation of $\hat H$ is diagonalized.

For TNS and related methods,
the simplest but not most efficient strategy of using \autoref{eq:SI_system} is to  use just one guess vector iteratively with $\ket{\tilde \Psi} = \ket{\Psi_{i-1}}$ and without keeping previously solved states, resulting in the inverse power method.\cite{Calculating2016rakhuba,ExcitedState2021baiardi} 
As an alternative,
one recent promising direction\cite{ExcitedState2021baiardi} of the shift-and-invert approach is to combine DMRG-like methods with the FEAST algorithm, which is based on contour integration,\cite{Densitymatrixbased2009polizzi,Contour2023brennan}
leading to multiple complex-valued $\sigma$ values in \autoref{eq:SI_system} and an additional numerical integration  of the basis states.
This, however, requires costly, complex algebra and the knowledge of how many eigenstates reside in a given energy interval, which is difficult to estimate for high-density-of-state regions.
Another method dubbed ``multiple shift block inverse iteration eigensolver'' is based on the canonical tensor decomposition and uses multiple real-valued $\sigma$ values and powers $P$ of $(\hat H - \sigma \hat I)^{-1}$. 
\cite{Computing2021kallullathil,Computing2023kallullathil}
The selection of the $\sigma$ values and the used maximum power $P$, however, is not directly clear. Hence, both this method and the DMRG-FEAST variant require multiple iterations to find the best parameters.

For the linear eigenvalue problem, perhaps the best-known usage of \autoref{eq:SI_system} is the shift-and-invert Lanczos method,\cite{Spectral1980ericsson,Templates2000bai} 
where $\ket{\tilde\Psi}$ is $\ket{\Psi_{i-1}}$.
Similar to the original Lanczos algorithm, the resulting basis
that spans a Krylov space 
of size $\Nkrylov$ consists of the polynomials $\{\ket{\Psi_i}\equiv(\hat H - \sigma \hat I)^{-(i-1)} \ket{\Psi_1} \}_{i=1}^{\Nkrylov}$.\cite{Iteration1950lanczos,Templates2000bai} 
By enforcing orthogonality of the polynomials using Gram-Schmidt-like orthogonalization, one obtains a simple three-term recursion for the polynomials. Then, $\hat H$ projected onto the space of orthogonal polynomials leads to a tridiagonal matrix that needs to be diagonalized to obtain approximate eigenvectors. 

While powerful for the linear eigensystem, the  shift-and-invert Lanczos method, however, requires  a very precise solution of the linear \autoref{eq:SI_system}, which is impractical for TNS methods.
In a paradigm shift, to avoid the cost of computing the linear equation accurately, in the inexact Lanczos method, the equation  is solved approximately. 
While this sounds trivial, it has profound consequences, such as $\hat H$ represented in the Lanczos basis turning from a tridiagonal into a dense matrix. 
The inexact approach can be understood as not trying to find the most optimal basis for diagonalization, which often is very costly,  but rather finding a basis that is good enough for finding eigenvalues and eigenvectors with a given target accuracy. 

The inexact Lanczos method has been used for various linear eigenvalue problems.\cite{New2000huang,Accelerating2001poirier,Preconditioned2002poirier,Filter2003sun,Preconditioned2020miao,Extended2013rewienski,Restarted2019dax}
For nonlinear TNS methods, however, 
only  the original Lanczos method has been used for solving ground states or Green's functions,\cite{Densitymatrix1995hallberg,Adaptive2011dargel,Lanczos2012dargel,RootN2022nocera,Improved2025wang,Inexact2025dektor}
and we are not aware that the inexact Lanczos method has been combined with TNSs. We note, however, that the aforementioned 
DMRG-FEAST\cite{ExcitedState2021baiardi} and the multiple shift block method\cite{Computing2021kallullathil,Computing2023kallullathil}
actually solve \autoref{eq:SI_system} inexactly as well. This motivates an extension of the inexact Lanczos method to TNSs.

In this work, we apply the inexact Lanczos method to the nonlinear eigenvalue problem using basis states consisting of TTNSs.
Vector algebra operations such as additions and operator multiplications can, in practice, only be achieved approximately for TTNSs. Consequently, this will require multiple adaptations to the original inexact Lanczos procedure. 
As a side product, we also present a simple way to find a good initial TTNS guess to initiate the Krylov space.
Using these adaptations, we apply our new method to three difficult test cases: (1) Finding 122 %
states for the 12-dimensional acetonitrile molecule, \ce{CH3CN}, which is a common and difficult benchmark system;\cite{Benchmarking2025larsson} 
(2) computing vibrational Fermi  resonance states for the 15-dimensional fluxional Zundel ion, \ce{H+.(H2O)2}, which are very difficult to retrieve;\cite{Vibrational2005hammer,Dynamics2007vendrell,Stateresolved2022larsson}
 and (3) computing, for the first time, full-dimensional excited eigenstates of the  33-dimensional, fluxional Eigen ion, \ce{H3O+.(H2O)3}, which is an extremely challenging, very correlated, and high-dimensional problem.\cite{HighLevel2017yu,Coupling2022schroder} 
As a by-product of our comparison, we also compute up to 1379 low-energy eigenstates using our previous DMRG-based TNS method for the Eigen ion.
The last two problems can currently be considered as one of the most difficult ones in vibrational quantum dynamics.
Given our results, we believe that our proposed methods will become a useful alternative to standard DMRG methods to compute excited states for very complicated systems such as high-dimensional fluxional molecules.
The proposed approach can also be applied to other fields, including DMRG-based electronic structure computations.\cite{Dynamical2002jeckelmann,Targeted2007dorando,TimeStep2017ronca,Minimal2020larsson,Density2020baiardi,Matrix2022larsson,LargeScale2022barcza,Block22023zhai,Simulating2024wahyutama}

The outline of this paper is as follows. 
\autoref{sec:theory} provides the details about the underlying theory of TTNSs, the required changes to the inexact Lanczos algorithm, and a new
way to efficiently compute initial guesses.
\autoref{sec:results} shows the performance of our new initial guess  and details our application to the aforementioned vibrational systems.
We conclude and provide our perspective on future developments in \autoref{sec:conclusions}.

\section{Theory}\label{sec:theory}
In this Section we briefly outline the theory of TTNSs (\autoref{sec:ttns}) and DMRG-methods to solve optimization problems (\autoref{sec:ttns_optimizations}).
We then discuss adjustments of the inexact Lanczos method needed to apply it to TTNSs (\autoref{sec:lanczos}), which is followed by a simple procedure to obtain the initial guess (\autoref{sec:opt_init_guess}).

\subsection{Tree tensor network states}\label{sec:ttns}

\begin{figure}[!htbp]
  \includegraphics[width=\columnwidth]{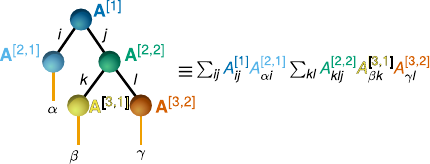}
    \caption{Example of a tree tensor network diagram and the corresponding contraction pattern. Compare with \autoref{eq:ttns}.}
  \label{fig:ttns}
\end{figure}

In the following, we give a brief introduction into the tree tensor network state (TTNS) approach. For an in-depth discussion and a comparison of ML-MCTDH, we refer the reader to our previous work;\cite{Computing2019larsson,Tensor2024larsson} 
see also \lits{Low2012chan,Practical2014orus,Tensor2021glaser}.
In the standard full configuration interaction approach, 
the $F$-dimensional wavefunction is expanded in terms of direct products of 
orthonormal ``primitive'' basis states  $\{\ket{\chi^{(f)}_{\alpha_f}}\}_{\alpha_f=1}^{N_f}$, where $N_f$ is the basis size and $f$ is the dimension where the basis is used. This leads to the following ansatz,
\begin{equation}
  \ket{\Psi} = \sum_{\alpha_1}^{N_1}  \sum_{\alpha_2}^{N_2} \dots \sum_{\alpha_F}^{N_F} %
  C_{\alpha_1 \alpha_2\dots \alpha_F} \bigotimes_{f}^{F} \ket{\chi^{(f)}_{\alpha_f}},
  \label{eq:fci}
\end{equation}
where all information is stored in one large  coefficient tensor $\matr C$.
For simplicity, throughout we assume that all tensors are real-valued.  
To alleviate the exponential scaling of the size of $\matr C$ with dimensionality $F$, 
TTNSs approximate $\matr C$ by a nested sum of products of smaller-dimensional tensors. 
For example, a three-dimensional tensor $\matr C$ could be approximated by a TTNS using
\begin{align}
  C_{\alpha \beta\gamma} &\approx 
   \sum_{ij} A^{[1]}_{ij} %
  A^{[2,1]}_{\alpha i} \sum_{kl} A^{[2,2]}_{klj} A^{[3,1]}_{\beta k} A^{[3,2]}_{\gamma l},\label{eq:ttns}
\end{align}
where the entries of the tensors $\matr A^{[\ell,h]}$ need to be determined. 
The sizes of $\matr A^{[\ell,h]}$ are called bond dimensions, $\bdim$, and the maximum bond dimension is $\bdimmax$.
The larger the $\bdimmax$, the  more accurate the approximation.
\autoref{eq:ttns} and similar equations can be more conveniently depicted using a diagrammatic notation, where each tensor is shown as a node, and each index is shown as a bond. Similar to Einstein's summation notation,\cite{Grundlage1916einstein} contraction/summation over common indices is represented by connecting the corresponding bonds of two nodes.
For example, \autoref{eq:ttns} as a diagram is shown in \autoref{fig:ttns}.
These diagrams are also useful for labeling each tensor. 
The labels  in $\matr A^{[\ell,h]}$ correspond to the layer $\ell$ and the horizontal position $h$ in that layer. We introduced this notation in \lit{Computing2019larsson}. It differs from common ML-MCTDH notation,\cite{Multilayer2008manthe,Tensor2024larsson} which is more powerful but not needed in this work. The single tensor (node) in layer $\ell=1$  is called the root tensor (root node).

We now discuss how the tensors can be orthogonalized, which is called canonicalization in the tensor network community. %
Inserting an invertible matrix $\matr T$ and its inverse into \autoref{eq:ttns} will not affect $\ket{\Psi}$, i.e., we can rewrite the right-hand side of \autoref{eq:ttns} as
\begin{equation}
\begin{split}
   \sum_{ijkl} A^{[1]}_{ij} %
   A^{[2,2]}_{klj} A^{[2,1]}_{\alpha i} A^{[3,1]}_{\beta k} A^{[3,2]}_{\gamma l} &=\\ %
   \sum_{ixyzkl} A^{[1]}_{ix}  %
   T_{xy} [T^{-1}]_{yz} %
   A^{[2,2]}_{klz} A^{[2,1]}_{\alpha i} A^{[3,1]}_{\beta k} A^{[3,2]}_{\gamma l}  %
\end{split}
  \label{eq:gaugeinv}
\end{equation}
This property leads to a gauge invariance and can be exploited in various ways. The most important exploitation is that $\matr T$ can be a unitary matrix that orthogonalizes one tensor, e.g., $\sum_{kl} A^{[2,2]}_{kli} A^{[2,2]}_{klj} = \delta_{ij}$. 
Combining the first two indices of $A^{[2,2]}_{klj}$  to one super-index $K$, $A^{[2,2]}_{Kj}$, then exposes the orthogonal matrix $\matr A^{[2,2]}$.
By convention, the tensors in the tree are always orthogonalized such that the corresponding matricized tensor has a column index that
connects the tensor in layer $\ell$ to one in layer $\ell-1$ (due to the tree structure, there will only be one index with this property)
and a row super-index that contains all other indices.
For example, $A^{[2,2]}_{klj}\equiv A^{[2,2]}_{Kj}$, which belongs to layer $2$, is connected through contraction of index $j$ to  $\matr A^{[1]}_{ij}$, which belongs to layer $1$.
The root tensor $\matr A^{[1]}_{ij}$ in the lowest layer $1$  is then viewed as a column vector.

Once the tensors are orthogonalized, we can  succinctly expand the wavefunction in terms of the root-node tensor $\matr A^{[1]}$ and high-dimensional configurations $\ket{\Phi^{[1]}_{ij}}$, which describe the remaining contraction with the primitive basis. For the decomposition in \autoref{eq:ttns} together with the primitive bases $\{\ket{\chi^{(1)}_\alpha}\}_\alpha$, $\{\ket{\chi^{(2)}_\beta}\}_\beta$, and $\{\ket{\chi^{(3)}_\gamma}\}_\gamma$, this gives
\begin{align}
  \ket{\Psi} =& \sum_{ij}  A^{[1]}_{ij} \ket{\Phi^{[1]}_{ij}}, \\
  \begin{split}
  \ket{\Phi^{[1]}_{ij}} =
  &A^{[2,1]}_{\alpha i}  \sum_{kl} A^{[2,2]}_{klj}  A^{[3,1]}_{\beta k} A^{[3,2]}_{\gamma l} \cdot\\%
  &\hphantom{\sum}\ket{\chi^{(1)}_\alpha}\otimes  \ket{\chi^{(2)}_\beta}\otimes  \ket{\chi^{(3)}_\gamma}.
  \end{split}
\label{eq:configs}
\end{align}

\begin{figure}[!htbp]
  \includegraphics[width=\columnwidth]{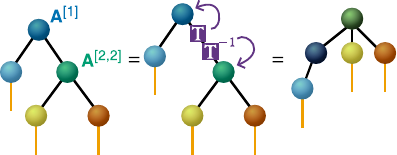}
    \caption{Example for change of root node. The unitary transformation matrix $\matr T$ orthogonalizes $\matr A^{[1]}$. The arrows indicate the tensors into which $\matr T$ and $\matr T^{-1}=\matr T^\dagger$ are absorbed. See the text for details.}
  \label{fig:ortho}
\end{figure}

The gauge invariance can not only be used to orthogonalize a TTNS in one given way, but it can also be used to change the root node in the tree, thus changing which indices in each tensor are interpreted as column index and which ones are interpreted as row super-index of an orthogonal matrix.
In the example in \autoref{eq:gaugeinv}, the transformation with $\matr T$ could make $A^{[1]}_{ix}$ orthogonal in the sense of $\sum_{i} A^{[1]}_{ix}  A^{[1]}_{iy} = \delta_{xy}$.  Then,
$A^{[1]}_{ix}$ and $A^{[2,2]}_{klj}$  exchange their place in the tree, turning $A^{[2,2]}_{klj}$ into the root node and moving $A^{[1]}_{ix}$ to the second layer.
This is shown diagrammatically  in  \autoref{fig:ortho}.
In practice, we use singular value decomposition (SVD) to perform the orthogonalization, which 
leads to natural basis functions that diagonalize one-dimensional reduced density matrices.
We also use this to dynamically adapt the bond dimension, essentially by discarding singular values below $\epsSVD$.

\subsection{Tree tensor network state optimizations}
\label{sec:ttns_optimizations}

\begin{figure}[!htbp]
  \includegraphics[width=.9\columnwidth]{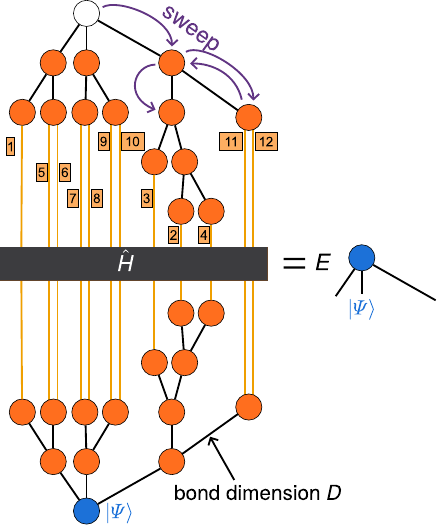}
    \caption{Effective eigenvalue problem, \autoref{eq:eigenvalue}, in tensor dagram notation. The gray box denotes the Hamiltonian, which is not shown in tensor form.
    The blue node corresponds to the eigenvector to be found. 
    The arrows indicate the first steps of the sweep procedure.
    The tree corresponds to that used for the \ce{CH3CN} simulations (see \autoref{sec:results} below), and the numbers in the orange boxes denote the mode ordering.
    Adapted from \lit{Benchmarking2025larsson}; licensed under a Creative Commons license.
    }
  \label{fig:sweep}
\end{figure}

The DMRG algorithm to compute ground states is based on a standard variational optimization, i.e.,\cite{Practical2014orus}
\begin{equation}
  \min_{\Psi} \bigl[\matrixe{\Psi}{\hat H}{\Psi} - E \braket{\Psi}{\Psi}\bigr],
\end{equation}
with $E$ being the energy and a Lagrange multiplier.
Instead of minimizing all parameters of $\ket{\Psi}$ concurrently, the DMRG algorithm is based on keeping all tensors but the root tensor fixed.
Using the previous example from \autoref{eq:ttns}
and the configurations, \autoref{eq:configs}, this then leads to a standard eigenvalue problem for the root tensor,
\begin{equation}
\sum_{kl} \matrixe{\Phi^{[1]}_{ij}}{\hat H}{\Phi^{[1]}_{kl}} A_{kl}^{[1]} = E A_{ij}^{[1]}\label{eq:eigenvalue}
\end{equation}
After solving \autoref{eq:eigenvalue}, $\ket{\Psi}$ is improved, and the TTNS is re-orthogonalized to a tensor connected to the previous root tensor. Then, an eigenvalue problem is solved for this new root tensor. By continuing this process for all tensors in the TTNS in one ``sweep,'' all tensors have been updated at least once.\cite{Density1992white,Densitymatrix1993white} These sweeps are repeated until convergence. The sweep procedure is called alternating least squares in the mathematics literature.\cite{Alternating2012holtz,Tensor2019hackbusch}
See \autoref{fig:sweep} for a sketch of this procedure in terms of diagrams and a larger TTNS.

The sweep algorithm not only applies to the variational ground state optimization but can also be used to fit $\ket{\Psi}$ to a sum of operators $\hat O_x$ applied to other states $\sum_x \hat O_x \ket{\tilde \psi_x}$ by solving\cite{Practical2014orus}
\begin{equation}
  \min_{\Psi} \| \ket{\Psi} - \sum_x \hat O_x \ket{\tilde\psi_x}\|^2.
  \label{eq:psi_fit}
\end{equation}
If $\hat O_x = \hat 1$, this corresponds to fitting a sum of TTNSs to one single TTNS. If the sum reduces to one TTNS $\ket{\tilde\psi}$, \autoref{eq:psi_fit} can be used to compress $\ket{\tilde\psi}$ to a new TTNS $\ket{\Psi}$ with a smaller bond dimension.
To solve \autoref{eq:psi_fit} in a sweep-like fashion,
we obtain the following update 
for the root tensor of $\ket{\Psi}$:
\begin{equation}
 A_{ij}^{[1]}  =\sum_{kl} \sum_x \matrixe{\Phi^{[1]}_{ij}}{\hat O_x}{\widetilde\Phi^{[1],x}_{kl}} \tilde A_{kl}^{[1],x},
 \label{eq:fitting}
\end{equation}
where $\widetilde\Phi^{[1],x}_{kl}$ ($\tilde A_{kl}^{[1],x}$) are the configurations (root tensors) of $\ket{\tilde \psi_x}$.
Note that alternatives to this fully variational optimization exist, such as summing up two TTNSs by stacking their tensors. Here, we will not consider these alternatives.

To solve linear systems of type  
\begin{equation}
\hat O \ket{\Psi} = \ket{\tilde \psi},\label{eq:lin_system}
\end{equation}
for positive definite $\hat O$, 
we can make use of\cite{Solution2012oseledets,Communication2014sharma}
\begin{equation}
  \min_{\Psi} \bigl[ \matrixe{\Psi}{\hat O}{\Psi}/2 - \braket{\Psi}{\tilde\psi}\bigr],
  \label{eq:min_lin_posdef}
\end{equation}
which, for the root tensor, leads to the update equation
\begin{equation}
 \sum_{kl} \matrixe{\Phi^{[1]}_{ij}}{\hat O}{\Phi^{[1]}_{kl}} A_{kl}^{[1]} = \sum_{kl}  \braket{\Phi^{[1]}_{ij}}{\widetilde\Phi^{[1]}_{kl}} \tilde A_{kl}^{[1]}.
 \label{eq:linear}
\end{equation}
For indefinite and positive semi-definite $\hat O$, \autoref{eq:min_lin_posdef} is not a proper functional, and instead, one has to consider 
\begin{equation}
  \min_{\Psi} \| \hat O \ket{\Psi} -  \ket{\tilde\psi}\|^2,
\end{equation}
which leads to normal equations containing $\hat O^2$.
To avoid the costly evaluation of $\hat O^2$, however,
here we follow previous strategies\cite{Dynamical2002jeckelmann,Solution2012oseledets,Calculating2016rakhuba,TimeStep2017ronca,ExcitedState2021baiardi}
and na\"ively use \autoref{eq:linear} even for indefinite $\hat O$. 
This then does not correspond to a minimization of a functional but instead corresponds to projecting \autoref{eq:lin_system}  onto the configurations of $\ket{\Psi}$ and $\ket{\tilde \psi}$. 
While mathematically, it is unclear why this should work, in practice, it has been shown to give a solution as long as a good initial guess of $\ket{\Psi}$ can be provided.\cite{Solution2012oseledets}

\subsection{Inexact Lanczos methods for tree tensor network states}\label{sec:lanczos}
Below, we discuss the adaptations to the inexact Lanczos method that are needed to make it work for TTNSs. 
This includes an approximate way to orthogonalize states (\autoref{sec:ortho}), 
an extension to the block Lanczos method, which allows us to simultaneously target  degenerate states  (\autoref{sec:block_lanczos}),
and a combination with a straightforward method to avoid recomputing previously computed states 
based on energy-shifting those states (\autoref{sec:shifting}).
Details on the implementation and an outline of the final algorithm are provided in \autoref{sec:implementation}.
\subsubsection{Approximate orthogonalization}
\label{sec:ortho}
Next to the approximate solution of \autoref{eq:SI_system}, the inexact Lanczos method applied to TTNSs requires that
every linear combination be approximate, as detailed in \autoref{sec:ttns_optimizations}. 
This has severe consequences, since orthogonalizing the Krylov states cannot be done exactly, unlike in the traditional inexact Lanczos method. 
An exact orthogonalization of one TTNS against another TTNS would require doubling the bond dimension.
Previous TNS-based methods that did not rely on solving the linear \autoref{eq:SI_system} avoided the explicit orthogonalization 
and instead used nonorthogonal or approximately orthogonal Krylov states and a subsequent orthogonalization of the final nonorthogonal Krylov space.\cite{Lanczos2012dargel,Chebyshev2021jiang,Improved2025wang}
However, we have found that, for our case, using nonorthogonal states quickly leads to an overcomplete basis and poor eigenvalue convergence, see also \lits{Inexact2025dektor,Improved2025wang} for a similar observation for the standard Lanczos approach combined with DMRG methods.
Here, instead, for each Krylov vector $\ket{\Psi_i}$, we use a modified Gram-Schmidt orthogonalization by performing $i-1$ individual TTNS fits, i.e., minimizing 
\begin{equation}
\left\|\ket{\Psi_i^\text{GS}} -  \left(\ket{\Psi_i} - \frac{\braket{\Psi_{k}}{\Psi_i}}{\braket{\Psi_i}{\Psi_i}} \ket{\Psi_{k}} \right) \right\|^2,
\end{equation}
where $k$ iterates from $1$ to $i-1$, and $\ket{\Psi_i^{\text{GS}}}$ replaces $\ket{\Psi_i}$ after each iteration.
This is done in the same way \autoref{eq:psi_fit} is solved.
Note that 
these fitting procedures are much faster than solving \autoref{eq:SI_system}, and the Krylov space size $\Nkrylov$ is kept low enough to avoid fitting too many states. 
Further, following the inexact approach, here we perform 
the orthogonalization  only approximately and use a fixed maximal bond dimension. Hence, our inexact Krylov vectors are still nonorthogonal, but their linear dependency is reduced, which dramatically improves the eigenvalue convergence.
\subsubsection{Dealing with near-degeneracies I: block Lanczos method}
\label{sec:block_lanczos}
Vibrational and vibronic eigenspectra have a very large density of states with many (quasi-)degeneracies in high-energy regions.\cite{Spectroscopic1981domcke,vibronic2024larsson,Benchmarking2025larsson}
This makes it difficult to resolve one eigenstate from others, even when using the shift-and-invert mode. Consequently, slow convergence and root flipping problems can occur.
To overcome this, one simple approach we use here is the block version of the Lanczos algorithm,\cite{Block1977golub} where $\Nblock$ orthogonal initial guesses are used, and for each guess, a separate Krylov  space is constructed. For a Krylov space of size $\Nkrylov$, we thus obtain a final space of size $\Nblock \cdot \Nkrylov$, which is then used for the diagonalization.
This approach works well if, in the spectrum close to $\sigma$, there are $N_\text{block}$ states that are well-separated from other states.\cite{Templates2000bai}
The addition of the block Lanczos method to our TTNS adaptation is trivial and only involves minor bookkeeping effort.

\subsubsection{Dealing with near-degeneracies II: State shifting}
\label{sec:shifting}
Different eigenstates have different convergence rates. To avoid further optimizing already converged states during Lanczos iterations, in different deflation methods, the converged states are removed from the Lanczos iteration, e.g., by orthogonalizing the Krylov vectors in subsequent Lanczos iterations against the converged states.
Such an orthogonalization is also important
for near-degenerate states where
changing $\sigma$ to target different eigenstates
in a separate Lanczos computation can lead to
states that have already been computed in previous Lanczos computations that used different $\sigma$ values.
Due to the approximate nature of TTNS orthogonalization, standard deflation methods work poorly in our TTNS adaptation.
Instead, we do not use deflation methods but 
resort to shifting the energy  $E_I$ of previously computed states $\ket{I}$,\cite{Iterative1973shavitt,CheMPS22014wouters,Computing2019larsson} which is our default way to compute thousands of low-energy states.\cite{Stateresolved2022larsson,vibronic2024larsson,Benchmarking2025larsson}
Importantly, we only need to shift those states whose energies are close to $\sigma$.
We shift the energy by modifying $\hat H$ according to 
\begin{equation}
  \hat H \to  \hat H + \sum_I (E_I + S) \ketbra{I}{I},
  \label{eq:H_shifting}
\end{equation} 
where $S$ is a large number that ensures energy separation.
\subsubsection{Implementation}
\label{sec:implementation}
The structure of the final algorithm is outlined in Algorithm~\ref{algo:lanczos}.
Note that, for simplicity, many input parameters and some other aspects, such as state shifting using \autoref{eq:H_shifting}, are not included explicitly.
Further, different bond dimensions and sweep schedules can be used for different TTNS operations. 
After creating new vectors, approximately orthogonalizing them and diagonalizing the current Krylov space in line \ref{line:diagonalization},  the eigenvectors and eigenenergies are sorted in line \ref{line:sorting}.
The sorting is done either by the deviation of the energies from $\sigma$, or by the user-provided function \texttt{sortingcriterion}, which allows for other sorting criteria, e.g., based on a physical property. 
Note that, such as in other iterative algorithms, the Hamiltonian and overlap matrices from previous Lanczos iterations can be re-used in subsequent iterations. 
The algorithm terminates if the first $\Nblock$ energies are converged according to a relative convergence criterion, $\tolkrylov$,
\begin{equation}
	\frac{\sum_{b=1}^{\Nblock}|E^{(b)}_{i+1}-E^{(b)}_{i}|}{\sum_{b=1}^{\Nblock}|E^{(b)}_{i+1}|} \le \tolkrylov
	\label{eq:errorcriterion}
\end{equation}
where $E_{i}$ and $E_{i+1}$ are the Lanczos block eigenvalues at iteration $i$ and $i+1$, respectively. 
This can be adapted to converge a larger number of states.
If the energies are not converged after the complete Krylov space is built up, the procedure is restarted using the first $\Nblock$ approximate eigenstates as an improved initial guess.
To gauge the accuracy of the approximate eigenstates, after final convergence, we compute the norm of the residual $\ket{r} = \hat H \ket{I} - E_I \ket{I}$
of all approximate eigenstates by approximately fitting $\ket{r}$ to $\hat H \ket{I} - E_I \ket{I}$,
which avoids a costly computation of $\hat H^2$.\cite{Lanczos2012dargel}
This fitting is only approximate, so the norm of the residual only serves as a very rough estimate of the accuracy of the energies.

\begin{figure}
\begin{algorithm}[H]
  \caption{Simplified pseudocode of the inexact Lanczos algorithm for TTNSs. See the text for details.}
  \label{algo:lanczos}
  \begin{algorithmic}[1]
    \Procedure{inexact\_lanczos}{
    $\{\ket{\Psi_1^{(b)}}\}_{b=1}^{N_\text{block}}$,
    $\Nkrylov$, $\sigma$, \texttt{sortingcriterion}}
    \State \texttt{basis} $\longleftarrow \{\ket{\Psi_1^{(b)}}\}_{b=1}^{N_\text{block}}$
      \For{$i=2, \dots, \Nkrylov$} %
        \For{$b=1,\dots, \Nblock$}
              \State Solve $(\hat H - \sigma \hat 1) \ket{\Psi_{i}^{(b)}} \approx \ket{\Psi_{i-1}^{(b)}}$
        \EndFor
        \For{$b=1,\dots, \Nblock$}
              \State Orthogonalize $\ket{\Psi_{i}^{(b)}}$ against  \texttt{basis} 
              \State  \texttt{basis} $\longleftarrow \text{\texttt{basis}} \cup \{ \ket{\Psi_{i}^{(b)}}\}$
        \EndFor
        \State Construct Hamiltonian matrix $\matr{H}$ using \texttt{basis}  
        \State Construct overlap matrix $\matr{S}$  using \texttt{basis}  
        \State Solve $\matr H \matr U = \matr S \matr U \matr E_{i}$
        \label{line:diagonalization} 
        \State Sort $\matr U$ and $\matr E_{i}$
        using \texttt{sortingcriterion}
        \label{line:sorting} 
        \State Get energies $\{E^{(b)}_{i}\}_{b=1}^{\Nblock}$ from $\matr E_{i}$ %
        \If{$\{E^{(b)}_{i}\}_{b=1}^{\Nblock}$  close to $\{E^{(b)}_{i-1}\}_{b=1}^{\Nblock}$}
            \State Exit loop.
        \EndIf
      \EndFor
            \State Construct approximate eigenstates:
            \For{$I=1,\dots,\text{len}(\texttt{basis})$}
                \State
                $\ket{I} \approx \sum_{x} U_{xI} \texttt{basis}_x$
        \label{line:construct_eigenstates} 
            \EndFor
    \EndProcedure
  \end{algorithmic}
\end{algorithm}
\end{figure}

The current algorithm can be improved in many ways, e.g., using improved restarting,\cite{Templates2000bai,ThickRestart2000wu,Variational2009matyus,Thickrestart2019shimizu,Computing1991morgana,Restarted2019dax} or refinements of the inexact solutions.\cite{Extended2013rewienski}
As these improvements need to be adapted to the TTNS case and might not work there, they are outside the scope of this work. 
For similar reasons, we have not attempted to use preconditioners that approximate the inverse of $\hat H - \sigma \hat 1$. 

While we do not use preconditioners in the overall Lanczos algorithm, we do use a Jacobi preconditioner for solving the update equation \autoref{eq:linear} during the sweeps together with the GCROT(m,k) algorithm,\cite{Simplified2010hicken}
as implemented in SciPy.\cite{SciPy2020virtanen} %
To gauge the convergence of the sweeps to solve  \autoref{eq:SI_system} using TTNSs, we monitor the energy of $\ket{\Psi_i}$ instead of the expression in  \autoref{eq:min_lin_posdef}, which is not a functional for indefinite operators.

\subsection{Optimizing the initial guess}
\label{sec:opt_init_guess}
Good initial guesses are crucial for the inexact Lanczos procedure. The need for good guesses can be alleviated by  starting with a random TTNS with a small bond dimension, including  Hartree product states, i.e., TTNSs with $\bdim=1$, and slowly growing the bond dimension after a few Lanczos iterations. 
However, we found it more efficient to start with a guess that is based on an unconverged DMRG optimization,  which itself does not rely on a good guess.
Thus, for the proposed procedure, we DMRG-optimize a random TTNS %
and abort the sweeps as soon as the energy drops below a target energy that is slightly larger than $\sigma$. At the node where the sweep aborts, we then re-solve the local eigenvalue problem, \autoref{eq:eigenvalue}, and target not the ground state of that node but an excited state close to the target energy using Davidson's algorithm.\cite{Iterative1975davidson}
For the block Lanczos guesses, we repeat this procedure using the state-shifted Hamiltonian, \autoref{eq:H_shifting}.

\section{Results and Discussion}
\label{sec:results}
In this Section, we apply the inexact Lanczos method using TTNSs to various challenging molecules.
To show the general convergence behavior, we first target a few states of \ce{CH3CN} using both initial DMRG-based guesses and random guesses (\autoref{sec:ch3cn_specific_state}).
This is followed by targeting the lowest 95 %
eigenstates and 
27 %
eigenstates in the interval from $2572$ to $\unit[2778]{\icm}$ %
of \ce{CH3CN} (\autoref{sec:ch3cn_all_states}). This includes protocols on how to perform multiple Lanczos runs to compute all states in one interval. 
We then compute Fermi resonance states of the Zundel ion (\autoref{sec:zundel}) and selected states of the Eigen ion (\autoref{sec:eigen}).

Throughout, the energies we show are the eigenvalues of the Hamiltonian represented in the TTNS Lanczos basis. That is, the energies belong to eigenvectors that are linear combinations of TTNSs. We do not show the less accurate energies obtained from the TTNSs that consist of the summation of TTNS basis states in line \ref{line:construct_eigenstates} in Algorithm \autoref{algo:lanczos}, as this summation can be done to arbitrary accuracy, if needed.  
In many cases, however, using the same $\bdimmax$ for the fitting as for the inexact Lanczos computations, and a relative convergence tolerance of $10^{-9}$, the fitting leads to energy changes that are lower than the eigenvalue error. 
To simplify the discussion, in the following, the range of $\hat H$ is zero-point-energy adjusted such that the ground state has an energy of $0$.

We now discuss some simulation parameters we used throughout, unless mentioned otherwise. 
We used a Krylov space dimension of $\Nkrylov=10$ with a relative eigenvalue convergence of $\tolkrylov=10^{-6}$. 
$\Nkrylov=10$ was chosen to avoid dealing with large sums of TTNSs, but the method also works for larger $\Nkrylov$ values.
To solve the linear system for each tensor, \autoref{eq:linear},
we used the  GCROT solver with an absolute (relative) convergence tolerance  of $10^{-9}$ ($10^{-2}$),
but the total number of GCROT iterations was capped at $10$ to minimize the runtime. 
Otherwise, default parameters were used. 
For state shifting (only applies to \ce{CH3CN}), we set the shift $S$ from  \autoref{eq:H_shifting} to $S=\unit[8499]{\icm}$.
For each sweep-based TTNS algorithm, we adapted the bond dimension 
using an SVD threshold of $\epsSVD=5\cdot 10^{-9}$ (see \autoref{sec:ttns}). To avoid bond dimension oscillations, after the bond dimension adaptation, we increased $\bdim$ to $\bdim +2$.\cite{Computing2019larsson} 
For the approximate orthogonalization (see \autoref{sec:ortho}), we used the same bond dimension parameters and a relative convergence tolerance of $10^{-2}$ with a maximum sweep  number of $40$.

\subsection{Initial benchmark: \ce{CH3CN}}
\label{sec:ch3cn_specific_state}
\ce{CH3CN} is a common benchmark problem, and we recently used it to compute up to 5000 eigenstates with an error estimate below $\unit[0.0007]{\icm}$.\cite{Benchmarking2025larsson}
Our used PES is based on the quartic force field from \lit{Calculations2005begue} in the parametrization from \lit{Using2011avila}, which is available in the supplementary information of \lit{Benchmarking2025larsson}.
We used the same tree structure as \lit{Benchmarking2025larsson}, which is depicted in \autoref{fig:sweep}.
The kinetic energy operator was based on the simplified $J=0$ Hamiltonian in normal coordinates, $- 0.5 \sum_\kappa \omega_\kappa\ \partial^2/ \partial \hat q_\kappa^2$, 
where $\hat q_\kappa$ is the position operator of mode $\kappa$ and $\omega_\kappa$ is its angular frequency.

We applied our method to target five different energy levels, which were either non-degenerate or degenerate. 
For this initial test, we used $\sigma$ values that are very close to reference energies. 
In Sections \ref{sec:ch3cn_all_states} and \ref{sec:zundel}, we will show that our method also works when $\sigma$ is not close to reference energies, which are not known in real-world problems.
The $\sigma$ values were the following: $360$ (state numbers 2, 3), $900$ (7), $1259$ (14, 15),
$1786$ (39), and $2066$ (57, 58) $\unit{\icm}$.
These values are selected based on different regions in the energy spectrum with different state densities. 
We used the block-Lanczos approach for the degenerate states.

Both DMRG and random guesses were based on $\bdimmax=3$.
We used a fixed number of $5$ sweeps to solve the linear system. %
See the Supplementary Material (SM) Table S2 for the target energies of the DMRG-based guess. 
For using this guess in the inexact Lanczos method, we immediately increased the maximum bond dimension to $\bdimmax=25$.
Instead,
for the random guess, we gradually increased $\bdimmax$ 
and tightened $\tolkrylov$
in three stages:
(1) $\bdimmax=3$ and $\tolkrylov=10^{-3}$; (2) $\bdimmax=10$ and $\tolkrylov=10^{-4}$; (3) $\bdimmax=25$ and $\tolkrylov=10^{-6}$.

\begin{figure}[!tbp]
        \centering
	\includegraphics[width=\columnwidth]{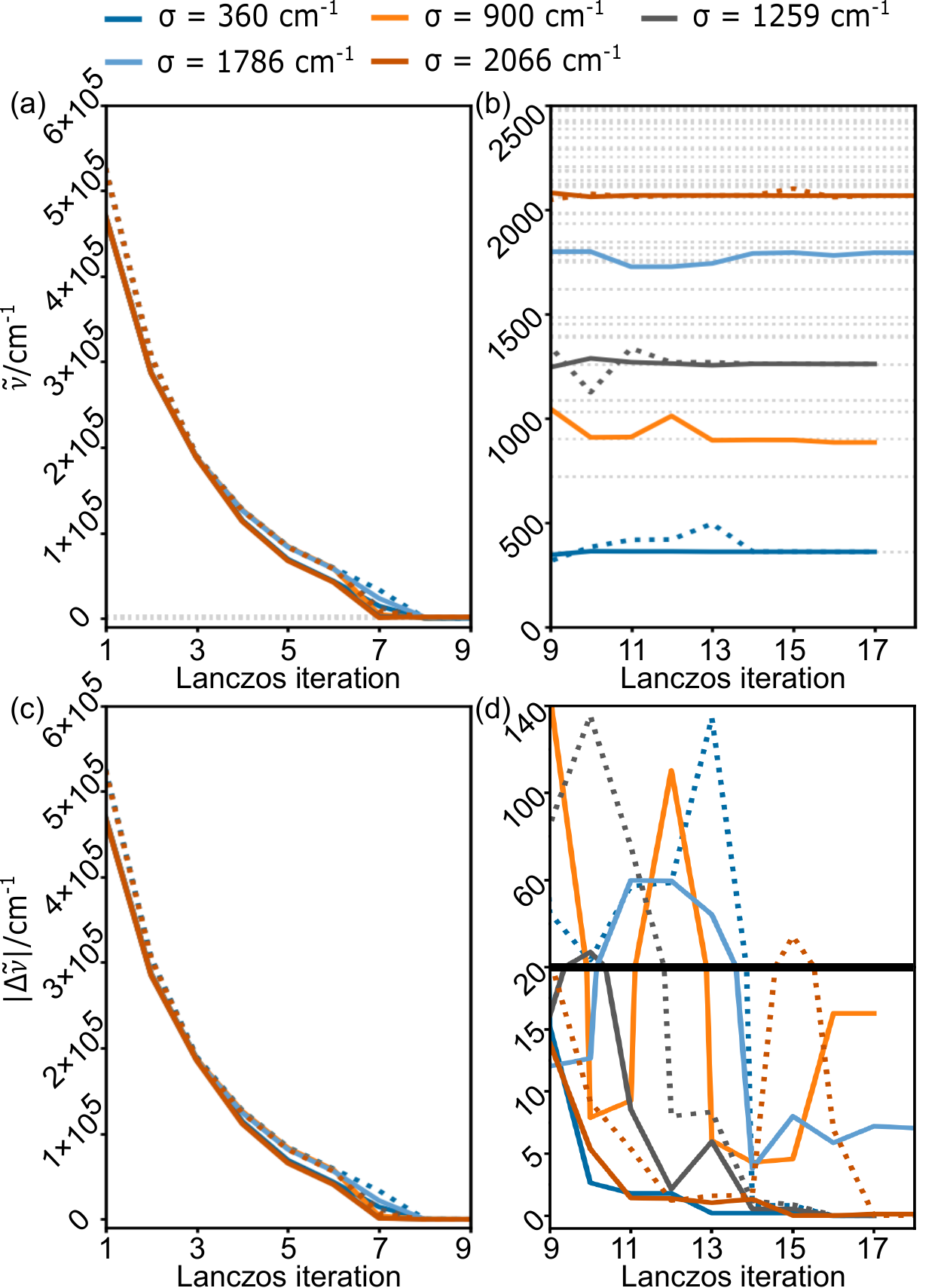}
\caption{Convergence of the inexact Lanczos methods for selected states of \ce{CH3CN} using a random initial guess.
Energy convergence as a function of cumulative iteration (a and b),
and the absolute value of the energy difference to the reference (c and d).
Due to different magnitudes, the convergence is split into two separate panels.
In (b),
the complete spectrum from the reference computations is shown as dotted gray lines.
Note further the different scales in (d), separated by a thick black horizontal line.
Iteration 1 contains the initial guess only. 
The computations are restarted after $\Nkrylov=10$ iterations.
For degenerate levels, the energies are shown as straight and dotted lines.
}
	\label{fig:ch3cn_specific_random}
\end{figure}

The convergence for the random guess is shown in \autoref{fig:ch3cn_specific_random},
which also displays our reference energies from \lit{Benchmarking2025larsson}.
To keep the comparison simple, here we only compare one single energy level for each target value $\sigma$, but there are other converged states close by.
Note that we show here cumulative iterations, which include restarts with larger bond dimensions for the random guess. 
The number of iterations for each stage is listed in the SM Table S1.
Importantly, throughout, we include the initial guess in the iteration count. 
For all targets, the random guesses require
8 to   14 iterations
to reach energy errors below $\unit[100]{\icm}$.
With the exception of one state,
all states converge within 17 or 18 iterations.
In contrast, the state for  $\sigma=\unit[1786]{\icm}$,
shows more erratic convergence behavior such as root-flipping, and requires 36 iterations to converge (not shown).
Despite the (premature) energy convergence of the Lanczos iterations, 
some of the final states exhibit large eigenvalue errors (\autoref{fig:ch3cn_specific_random}d).
This is also evident from the large residual norms (not shown) of the states with large eigenvalue error.
We will use residual norms as an additional convergence measure in \autoref{sec:ch3cn_all_states}.

\begin{figure}[!tbp]
        \centering
	\includegraphics[width=\columnwidth]{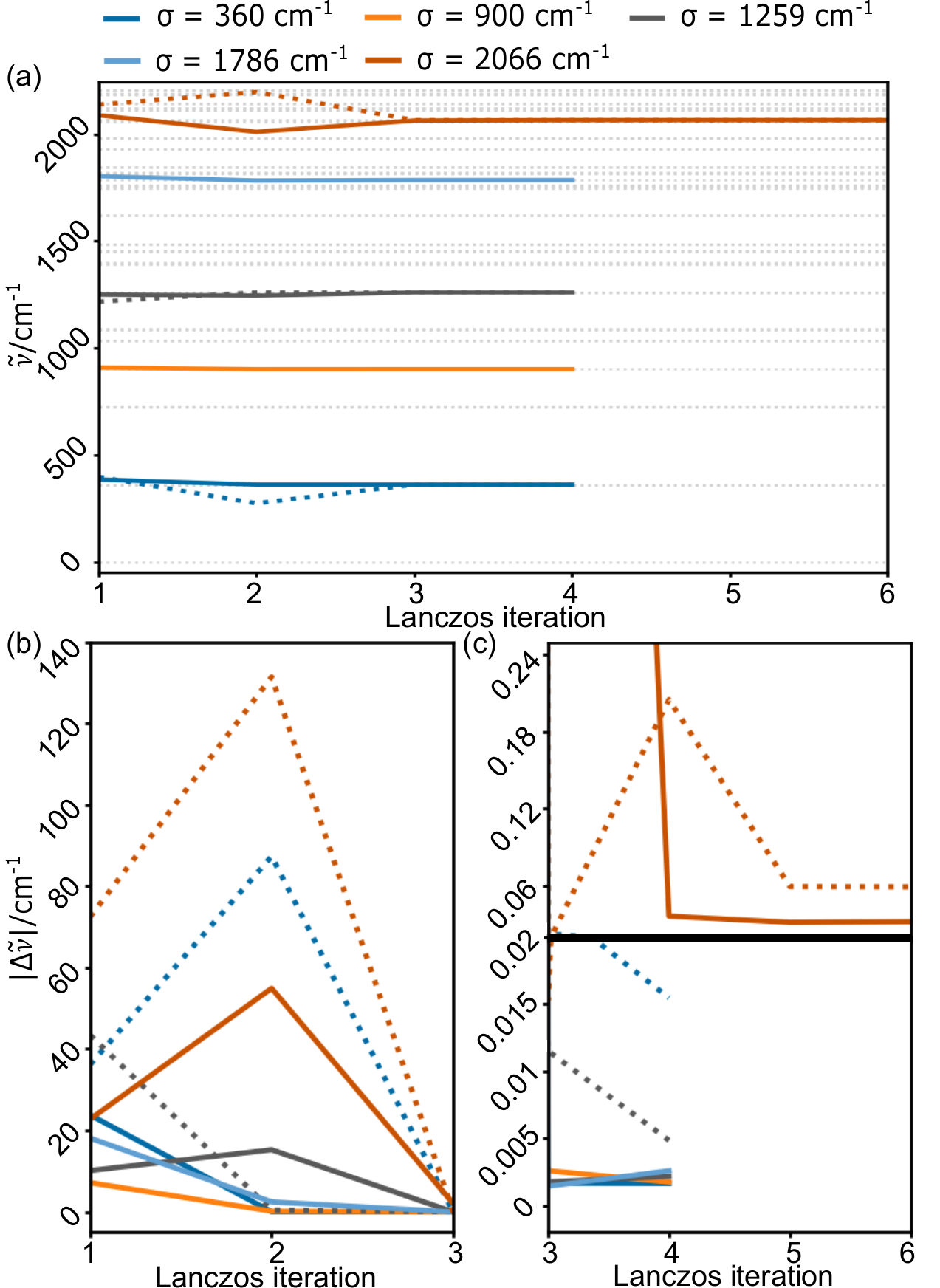}
\caption{Same as \autoref{fig:ch3cn_specific_random} but for a DMRG-based initial guess.}
	\label{fig:ch3cn_specific_dmrg}
\end{figure}

The convergence dramatically changes once the DMRG guess is used, as shown in \autoref{fig:ch3cn_specific_dmrg}.
Selecting the energy with the lowest error for degenerate pairs,
four out of the five states converge after just 4
iterations with an error below $\unit[0.005]{\icm}$. The state for $\sigma=\unit[2066]{\icm}$ converges after 6 iterations with an error below $\unit[0.06]{\icm}$. 
Compared to the random guess, 
the fast convergence is also reflected in the runtimes, as 
the DMRG-based guesses lead to a runtime that often is two orders of magnitude faster, even though the random guess computations start with smaller bond dimensions. 
In both cases, however, it can happen that during the sweep-based optimization of \autoref{eq:SI_system}, the solution oscillates or diverges after some iterations. Particularly, this can happen in the first Lanczos iteration, when the initial guess is poor and solutions are accurate to only $\unit[\sim 10]{\icm}$. In the following iterations, convergence frequently improves to $\unit[\sim0.1]{\icm}$. 
For these reasons, it is important to fix the total number of sweeps to low values of $\mathcal O(10)$. %
This works in most cases and gives reasonable approximations of \autoref{eq:SI_system}.

\subsection{Computing all states within an energy interval: \ce{CH3CN}}
\label{sec:ch3cn_all_states}
A common scenario is to compute all eigenstates within a given energy window. To demonstrate that this is possible with the TTNS inexact Lanczos approach, we compute two different energy regions. The first region covers all %
95 %
low-energy eigenstates up to an excitation energy of
$\unit[2346]{\icm}$. %
The second region covers  %
27 %
states between 
$\unit[2572]{\icm}$ and $\unit[2778]{\icm}$. %
The reason for this particular energy interval is that 
our previous high-accuracy benchmark on \ce{CH3CN} revealed that some eigenstate methods have difficulties in converging this region.\cite{Benchmarking2025larsson}

To achieve the computation of all states within an energy interval, we not only need multiple inexact Lanczos computations, but  we also need to ensure that we do not miss eigenstates. We approached these needs by using two stages based on TTNSs with different bond dimensions in each stage. The first stage is about ensuring that all states within an interval are computed using low-$\bdimmax$ TTNSs. These TTNSs are then refined/re-optimized using larger $\bdimmax$ values in the second stage. The refinement is repeated by successively increasing $\bdimmax$ until the required accuracy is achieved. 
The two stages are described below. 
Note that we did not strictly follow the parameters mentioned below, 
as the proposed procedure is not fully black-box. Additional details and a few exceptions with slightly different parameters are found in the SM Section S2.

\subsubsection{Finding all states in one interval}
\label{sec:ch3cn_all_states_in_one_interval}

To initialize our procedure to find all states, we set $\sigma$ to the lowest energy of the interval. From each inexact Lanczos computation, we considered not only the eigenstate close to $\sigma$ but also all  states
where the summed squared overlap to already computed eigenstates was less than $0.06$ %
and where %
the norm of the eigenstate residual was less than  a value ranging from $95$ to $\unit[150]{\icm}$. 
Note that the residual computation is approximate (see \autoref{sec:implementation}), so this 
heuristic parameter does not reflect the actual accuracy, which is much higher.
To double-check that no eigenstates have been missed and to retrieve degenerate eigenpairs, in the next computation we set $\sigma$ to the lowest eigenvalue of the previously retrieved eigenstates and shift those in energy (see below).
As new TTNS guess, we used one of the approximate eigenstates with a large residual, typically the one closest in energy to the target. %
We repeated this until no additional states had been found, and then performed an additional check,
where we again set $\sigma$ to the lowest eigenvalue of previously found states
but used a DMRG-based guess.
In some cases, this is important to find states belonging to different irreps.
Once we were sure that no eigenstate was missed at one $\sigma$ value, we increased $\sigma$ to the smallest energy of the approximate states with large residual norm that have not yet been added to our list of eigenstates. 
We used the corresponding TTNS as the initial guess for the next computation.
In addition, to avoid converging to previously found states, in all Lanczos optimizations, we shifted a small number of %
previously computed eigenstates closest to $\sigma$.
The shifting was also used for generating the  DMRG-based guess.
For the low-energy (high-energy) region, we shifted 10 (20) states.
For states with higher energies, which are more complex than those with smaller energies,\cite{Benchmarking2025larsson}
we needed to tighten the parameters. 
The linear solver parameters used for different eigenstates are shown in \autoref{tab:ch3cn_parameters}.

We iterated the described procedure until all eigenstates were retrieved.
While this requires repeated computations around the same target energy, the computational cost is modest as the bond dimension can  be set to low values.
Specifically, we used $\bdimmax=10$ for the first (low-energy) energy interval,
 and $\bdimmax=20$  for the second (high-energy) energy interval. %

\begin{table}[!htbp]
	\caption{Linear solver parameters for computing all states in an energy interval for \ce{CH3CN}. 
	For the solver GCROT(m,k) and for the DMRG-like sweeps, the relative tolerance is shown, next to the max.~number of iterations/sweeps in parentheses.
	}
\label{tab:ch3cn_parameters}
	\begin{tabular}{l@{\hskip 10pt}c@{\hskip 10pt}c@{\hskip 10pt}c@{\hskip 10pt}c}
	\toprule
	States & \multicolumn{2}{c}{$\bdimmax=10/20$} & \multicolumn{2}{c}{$\bdimmax=30/40$} \\
	 & GCROT &sweep &  GCROT &sweep \\
	\midrule
	Low   &         \\
		1-58  & $10^{-2} (10)$ & $0.05 (5)$ & $10^{-2} (10)$  &$0.05 (5)$       \\
		59-71 & $10^{-2} (10)$ & $0.05 (5)$ & $10^{-4} (20)$  &$0 (5)$ \\
	72-95 & $10^{-2} (10)$ & $0 (5)$  &  $10^{-4} (20)$  &$0 (5)$ \\
\midrule
High & \\
142-168 &  $10^{-4} (20)$&$0 (5)$  & $10^{-4} (20)$&$0 (5)$   \\
\bottomrule
\end{tabular}
\end{table}

\subsubsection{State refining}
\label{sec:ch3cn_state_refinement}
After the retrieval of the complete list of approximate eigenstates, 
we increased the max.~bond dimension and refined the states using the parameters shown in \autoref{tab:ch3cn_parameters}. 
To deal with near-degeneracies, here we used the block inexact Lanczos approach.\footnote{For a few outliers that did not converge, we used state shifting instread, see the SM Section S2.B for details.} 
We grouped all eigenstates based on how energetically close they are and used them as an initial guess, setting $\sigma$ to the median of the eigenvalues in one block.
We used this block refinement repeatedly, and, for each iteration, we increased the bond dimension in increments of $10$ until $\bdimmax=40$ was reached.
We used smaller energy intervals and thus smaller block sizes for states with larger bond dimensions, because these are more accurate and, thus, expose degeneracies well.
In some cases where we noticed larger overlaps to previously computed states or large energy differences compared to previous refinements with smaller $\bdimmax$ values, we merged blocks and repeated the optimization.
In all cases, 
the maximum block size did not exceed $\Nblock=8$.
\subsubsection{Results}
\begin{figure}[!tbp]
\includegraphics[width=\columnwidth]{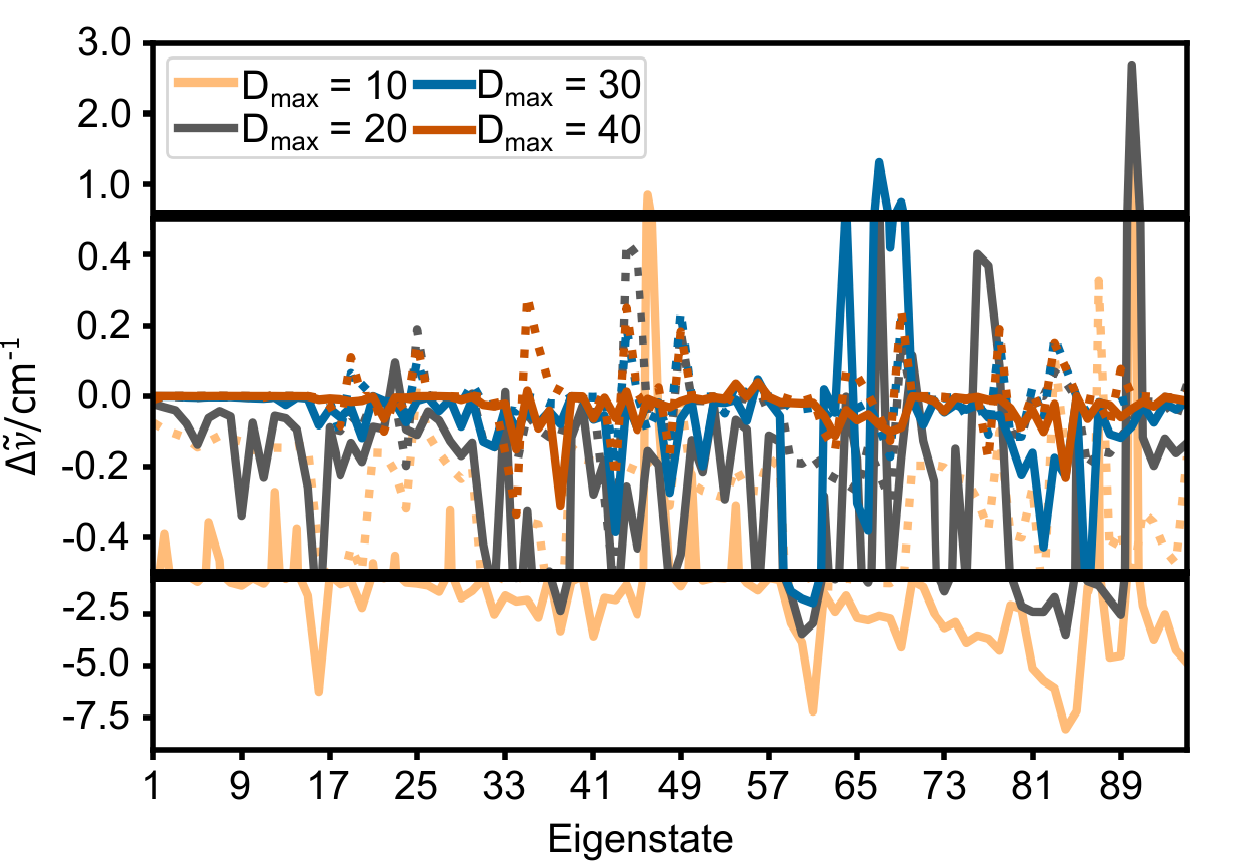}
	\caption{\ce{CH3CN} energy error for inexact Lanczos TTNSs  (lines) in comparison to DMRG-based TTNS energy errors (dots) for different max.~bond dimensions. Shown is the low-energy interval. Note the three different ordinate scales separated by two thick black horizontal lines.}
	\label{fig:ch3cn_all_low}
\end{figure}

Using our  benchmark energies from \lit{Benchmarking2025larsson} as reference,
the energy errors of the TTNS inexact Lanczos computations for the low-energy interval are depicted %
in \autoref{fig:ch3cn_all_low} for different bond dimensions. 
The energy error is defined as $\Delta \tilde \nu=\tilde \nu_\text{reference} - \tilde \nu_\text{inex.~Lanczos}$.
To compare these energies with TTNS DMRG energies, we further performed DMRG eigenstate computations using our  state-shifting approach for each $\bdimmax$ and with similar convergence parameters.
Note that these computations were different from \lit{Computing2019larsson}, where we used a combination of state-shifting and state-averaging.

The largest and the average absolute errors are listed in \autoref{tab:ch3cn_errors}. 
Note that the maximum error is dominated by outliers, but the general trend is similar to that of the maximum and average errors, as shown in \autoref{fig:ch3cn_all_low}.
Since the initial $\bdim=10$ inexact Lanczos computations were not designed to achieve high accuracy, their errors are significantly larger (a max.~error of $\unit[8.1]{\icm}$) than the corresponding DMRG ones (a max.~error of $\unit[1.5]{\icm}$).
This changes once we refine the states, which systematically decreases the error. 
The final $\bdimmax=40$ inexact Lanczos errors are similar to the DMRG errors, and, in fact, are slightly smaller.
However, the used relative energy convergence tolerance of $10^{-6}$ limits the accuracy for both the inexact Lanczos and the DMRG energies, which is seen by an \emph{increase} of the DMRG errors from  $\bdimmax=30$ to $40$. Tightening the convergence tolerance again decreases the error.
Hence, the final $\bdimmax=40$ inexact Lanczos energies can be regarded as fully converged to the used convergence tolerance.

\begin{table}[!htbp]
\centering
	\caption{TTNS inexact Lanczos and DMRG maximum and averaged absolute errors in $\unit{\icm}$ for different max.~bond dimensions in both energy intervals.}
	\label{tab:ch3cn_errors}
\begin{tabular}{l@{\hskip 10pt}c@{\hskip 10pt}d@{\hskip 10pt}d@{\hskip 10pt}d@{\hskip 10pt}d}
\toprule
	Interval & $\bdimmax$ &  \multicolumn{2}{c}{$|\Delta\tilde\nu_{I,\text{Lanczos}}|$} & \multicolumn{2}{c}{$|\Delta\tilde\nu_{I,\text{DMRG}}|$} \\
	& &  \multicolumn{1}{c}{$\max$.} & \multicolumn{1}{c}{$\text{av.}$} &  \multicolumn{1}{c}{$\max$.} & \multicolumn{1}{c}{$\text{av.}$}\\
\midrule
	Low  & 10 & 8.1   & 2.1    & 1.5   & 0.36 \\
	Low  & 20 & 3.5   & 0.55   & 0.56  & 0.082  \\
	Low  & 30 & 2.0   & 0.15   & 0.24  & 0.037 \\
	Low  & 40 & 0.31  & 0.030   & 0.34  & 0.047  \\
	High & 20 & 11  & 1.8   & 0.52  & 0.17 \\
	High & 30 & 0.96  & 0.27   & 0.24  & 0.070 \\
	High & 40 & 0.35  & 0.078  & 0.30  & 0.069  \\
\bottomrule
\end{tabular}
\end{table}

\begin{figure}[!tbp]
	\includegraphics[width=\columnwidth]{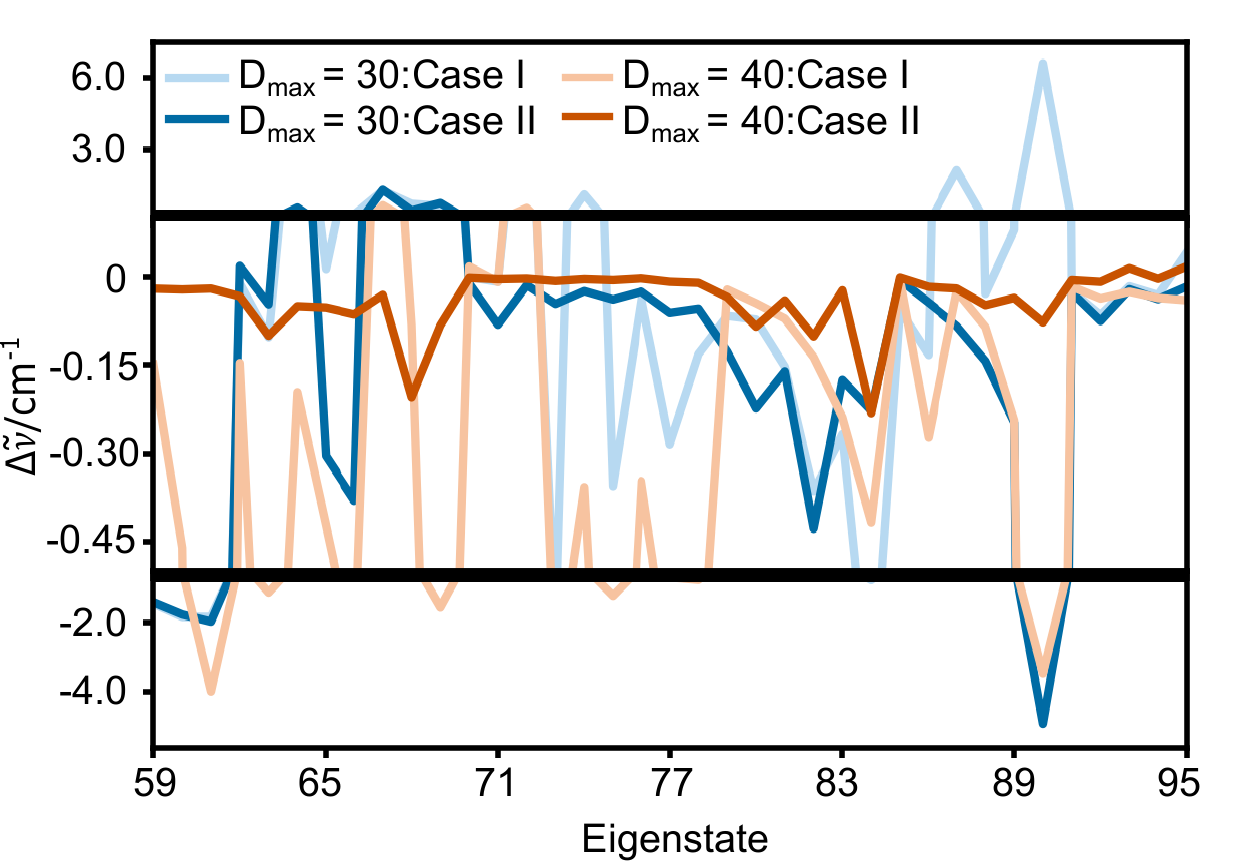}
	\caption{\ce{CH3CN} energy errors with linear solver parameters that are loose (case I, pale colors) compared to ones that are tight (case II, dark colors).
	Case I (II) corresponds to the $\bdimmax=10/20$ ($30/40$) parameters from  \autoref{tab:ch3cn_parameters}.
	For better comparison, the shown errors correspond to the simulations without a re-optimization using a combination of blocks as described in \autoref{sec:ch3cn_state_refinement}. The ``case II'' errors thus differ slightly from those shown in  \autoref{fig:ch3cn_all_low}.
 Note the three different ordinate scales separated by two thick black horizontal lines.}
        \label{fig:ch3cn_param_comparison}
\end{figure}

For $\bdimmax \ge 30$, we used tighter linear solver convergence parameters, as shown in \autoref{tab:ch3cn_parameters}. 
A comparison to results based on looser parameters is shown in \autoref{fig:ch3cn_param_comparison}.
With tightened parameters, for $\bdimmax=30$, the maximum (average) error decreases by $\unit[1.7]{\icm}$ ($\unit[0.16]{\icm}$). %
The effect is more dramatic for $\bdimmax=40$, where 
the maximum (average) error decreases by $\unit[3.8]{\icm}$ ($\unit[0.48]{\icm}$).
This indicates that, while the inexact Lanczos procedure is relatively robust with respect to the accuracy of the solutions of the linear system, \autoref{eq:SI_system},
for high target accuracies, the solver parameters cannot be too loose, and the energy convergence needs to be tested. In our case, this is particularly the GCROT convergence, whereas the max.~number of used DMRG-like sweeps can be set to relatively low values (5 in this case).

\begin{figure}[!tbp]
\includegraphics[width=\columnwidth]{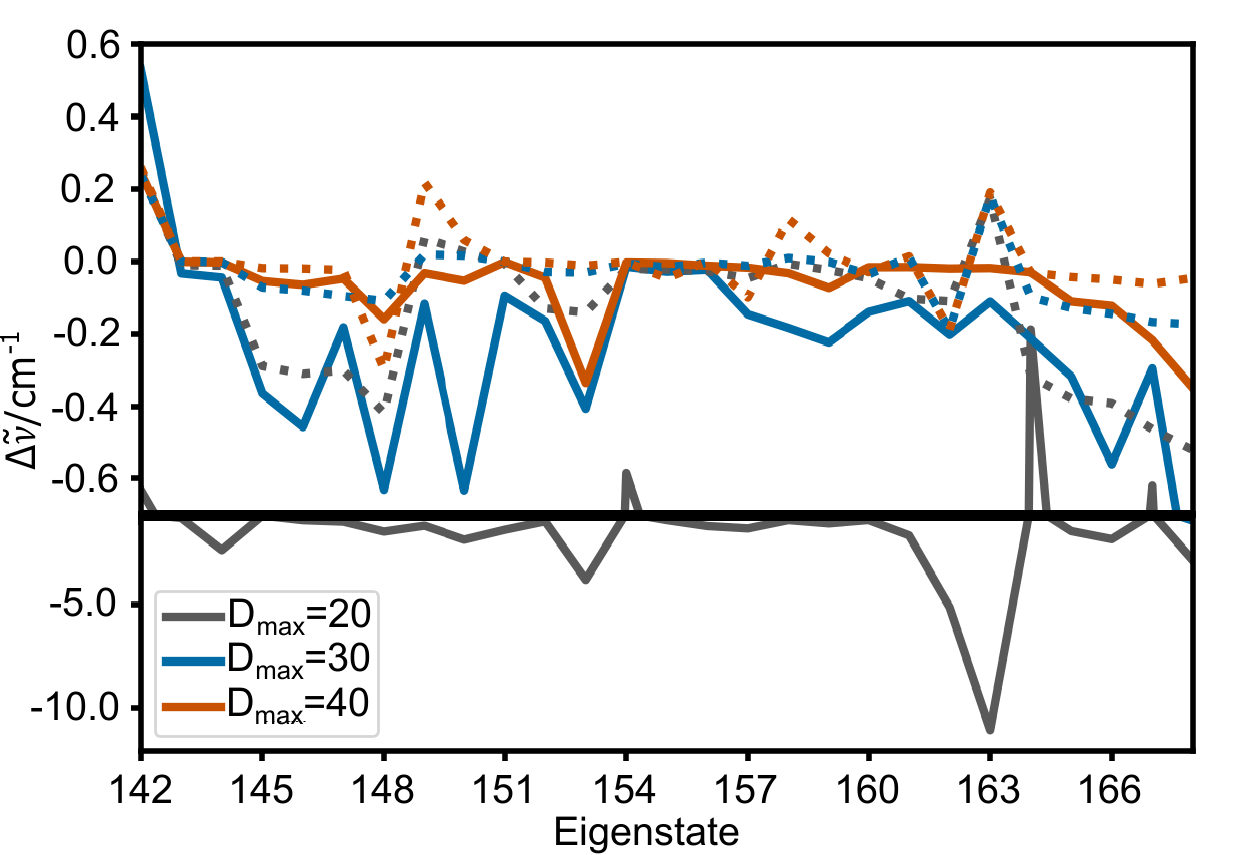}
\caption{Same as \autoref{fig:ch3cn_all_low} but for the high-energy interval.}
	\label{fig:ch3cn_all_high}
\end{figure}

The comparison of the high-energy interval is displayed in \autoref{fig:ch3cn_all_high}.  Like the low-energy interval, the initial Lanczos computation for $\bdimmax=20$ leads to relatively large errors, which are then significantly improved in the subsequent $\bdimmax=30$ and $\bdimmax=40$ refinements.
The final $\bdimmax=40$ energy errors are similar to the DMRG errors.

Note that not all DMRG and the Lanczos energies are fully variational, which is due to spurious overlaps between states. 
We recently resolved this issue by  an additional diagonalization of $\hat H$ in the bases of the TTNS states using a generalized eigenvalue problem, which increases the accuracy.\cite{Benchmarking2025larsson}
While this mostly restores variationality, %
this also increases 
the energy error of the Lanczos-optimized states, which is in contrast to the DMRG-optimized states.  We speculate that this is due to noise present in the Lanczos-optimized states. %
See the SM Section S2.D for more details.
\subsection{Fermi resonance in the Zundel ion}
\label{sec:zundel}
A dominant doublet in the IR spectrum of the Zundel ion ($\ce{(H2O)2H+}$) around $\tilde \nu = \unit[1000]{\icm}$ has intrigued scientists for many years,\cite{Vibrational2005hammer,Vibrational2006kaledin,Dynamics2007vendrell,Full2007vendrell,How2014rossi,Reduced2019bertaina,Stateresolved2022larsson,Revisiting2025ma}
before full-dimensional MCTDH simulations could reveal the origin -- a Fermi resonance between a state with two quanta in the wagging motion and one quantum in the \ce{O}-\ce{O}-stretch motion, and another state with one quantum in the proton transfer motion.\cite{Dynamics2007vendrell,Dynamics2007vendrell}
The assignment, however, was very tedious as no eigenstates but wavepacket propagations were available. With TTNS computations we recently computed 1000 states and revealed more subtle resonance effects.\cite{Stateresolved2022larsson} Computing more than 120 states was necessary to reach the two eigenstates associated with the doublet.

Here, we show that our new inexact Lanczos approach is able to target these eigenstates \emph{directly}, even with a poor initial guess. 
Our computations were based on the setup from \lit{Stateresolved2022larsson}, which uses the
sum-of-product approximation\cite{Transforming2020schroder}
of the PES from \lit{Automated2019schran},
and the polyspherical coordinates from \lit{Full2009vendrell}.
For the inexact Lanczos computations, we set $\bdimmax=50$ with $\epsSVD=10^{-6}$,
a GCROT relative tolerance of $10^{-4}$ ($20$ iterations maximum), and a fixed number of $7$ sweeps for the linear TTNS solver. %
The relative convergence tolerance for the orthogonalization was set to $10^{-8}$ ($40$ iterations maximum).
The inexact Lanczos computations used $\sigma=\unit[1000]{\icm}$.
For the final TTNS linear combination to plot the states, we used $\bdimmax=100$.
To compare with  states of similar accuracy,
we performed reference TTNS DMRG computations using the same $\bdimmax=50$. %

We created our initial state by applying the coordinate operator in the proton shuttle mode onto the ground state.
Such a state has previously been used to assign the doublet.\cite{Full2007vendrell,Stateresolved2022larsson}
Importantly, this initial state does not include any excitations in the wagging or \ce{O}-\ce{O}-stretch motion, which are crucial for an appropriate description of the doublet. We chose this guess specifically for this reason, as only a proton-transfer excitation in this region of the spectrum was found in unconverged vibrational configuration interaction computations,\cite{Vibrational2005hammer}
but not a contribution of the wagging motions.

\begin{figure}[!htbp]
  \includegraphics[width=\columnwidth]{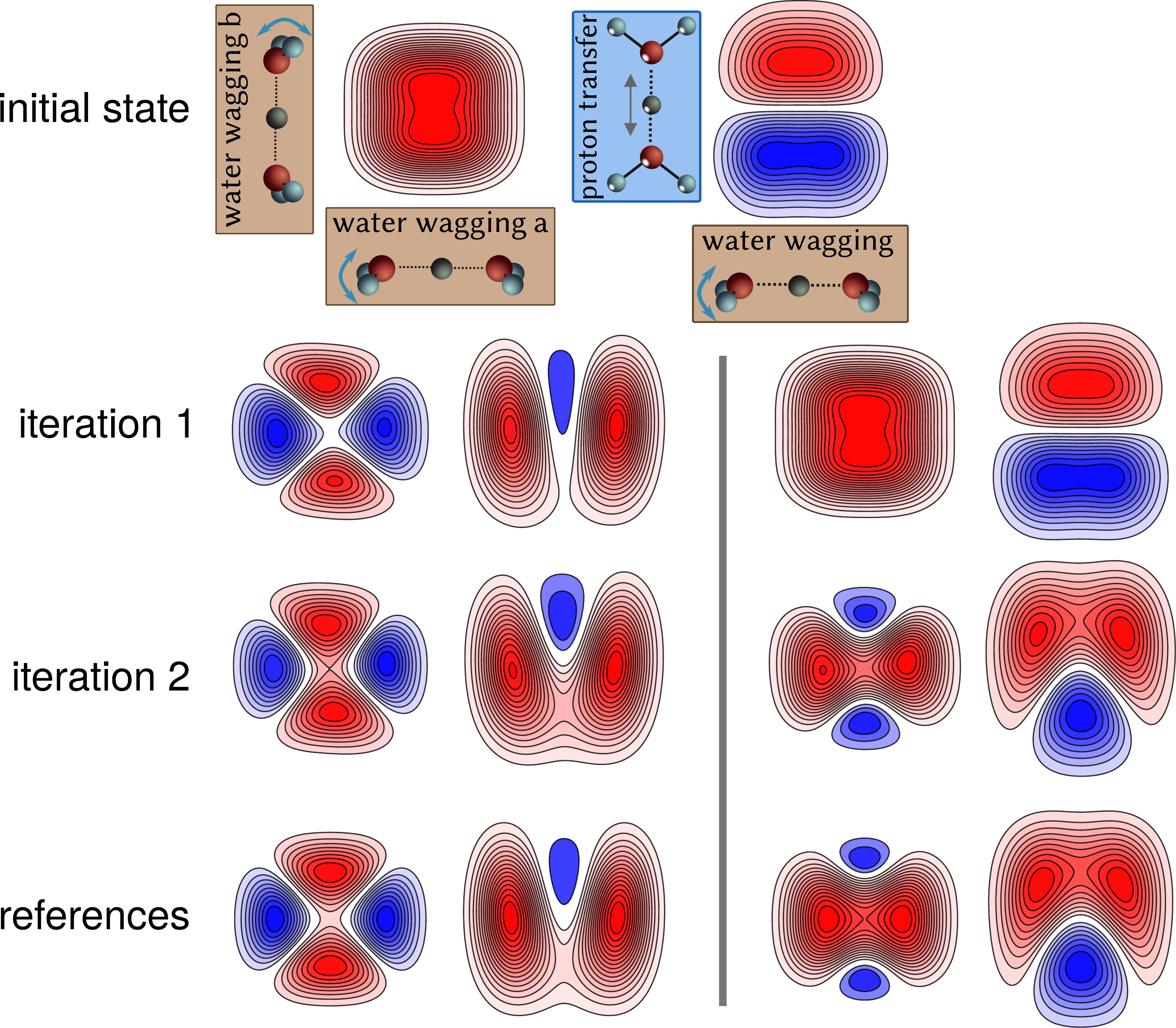}
    \caption{Inexact Lanczos optimization for the Fermi resonance in the Zundel ion around $\tilde \nu=\unit[1000]{cm^{-1}}$. 
    The first row shows the wavefunction cuts of the initial guess state  along the two water wagging coordinates, and along the proton transfer and one of the water wagging coordinates. 
    The second and third rows display the cuts of the resulting eigenstates in iterations 1 and 2 (iteration 2 contains three basis states and only the targeted ones are shown). The last row shows the DMRG references. %
    Red (blue) regions correspond to the positive (negative) wavefunction values. 
    }
  \label{fig:zundel}
\end{figure}

\autoref{fig:zundel} shows the wavefunction cuts (using the procedure from \lit{Stateresolved2022larsson}) of the initial guess state, and the important states at the first two Lanczos iterations in comparison with TTNS DMRG references (the iteration count does not include the guess).
Surprisingly, already in the first iteration, e.g., after one application of $(\hat H - \sigma)^{-1}$ onto the initial state, a state with a squared overlap of $0.68$ onto one of the reference states is obtained. This state shows all qualitative features of the Fermi resonance reference state, including the two excitations in the wagging motions that do not exist in the initial state.
The second Lanczos iteration improves this state further, leading to a squared overlap with the reference state of $0.90$.
In addition, in the second iteration the second eigenstate that dominates the doublet in the IR spectrum appears with a squared overlap to the reference of $0.88$.
Thus, just two applications of $(\hat H - \sigma)^{-1}$ onto the initial state suffice to reproduce the states contributing to the doublet.

Even though two iterations suffice to reach qualitative convergence, 
the states are not quantitatively converged, 
and the energy of the Lanczos eigenstates deviates by  $\unit[20]{\icm}$, compared to the TTNS DMRG references.\footnote{Note that the DMRG references with $D_\text{max}=50$ are converged to $\sim\unit[5]{\icm}$, compared to more accurate $D_\text{max}=150$ values from \lit{Stateresolved2022larsson}. We chose the  $D_\text{max}=50$  references as this value was also used in the Lanczos procedure.}
While in subsequent Lanczos iterations (not shown), the energy error improved, and the squared overlap to the references \emph{decreased} to $0.7$ at iteration 14. This indicates that the Lanczos basis might not be flexible enough, and the variational procedure  improves the energy at the price of deteriorating the wavefunction, which is a classical dilemma of variational optimization.\cite{Molecular2013helgaker}
While we could have used our procedure from \autoref{sec:ch3cn_all_states} to improve the states to reach a better accuracy, here, we refrain from it, as the purpose of this demonstration is to show that our eigenstate procedure allows us to quickly obtain qualitatively good solutions even with initial guesses that lack key characteristics of the targeted states.

\subsection{Individual state targeting: Eigen ion}
\label{sec:eigen}

The 33-dimensional Eigen ion is both fluxional and highly correlated,\cite{Coupling2022schroder,Tensor2024larsson} 
and even optimizing the ground state is challenging.\cite{Tensor2024larsson}
Despite this,
here, we demonstrate that TTNS inexact Lanczos and DMRG computations are still possible for the Eigen ion.
We used the Hamiltonian from \lit{Coupling2022schroder} and our optimized\cite{Computing2019larsson} tree structure from \lit{Tensor2024larsson}.
To have a reference to compare to, we first performed two separate TTNS DMRG computations using $\bdimmax=70$ and $150$, and diagonalized the final states. For $\bdimmax=70$ ($150$), we computed the lowest 1379  (237) eigenstates up to an excitation energy of $\unit[\sim350]{\icm}$ ($\unit[\sim190]{\icm}$).
More details on these computations are found in the SM Section S3, and a detailed analysis of these states will be published elsewhere.
The non-diagonalized (diagonalized) DMRG $\bdimmax=70$ energies have a maximum absolute deviation from the $\bdimmax=150$ energies of $\unit[1.7]{\icm}$ ($\unit[6.3]{\icm}$).

For the TTNS inexact Lanczos computations, we used   $\bdimmax=70$ and
computed three eigenstates with target values of $\sigma=\unit[1.22]{\icm}$, $\unit[79.22]{\icm}$, and $\unit[172.91]{\icm}$, respectively. 
These values were chosen to be close but not identical to $\bdimmax=70$ DMRG reference values, and the number of digits used was chosen to obtain integer values after adding the zero-point energy to $\sigma$.
The first value is close to the ground and first excited states,
the second value is close to state 39 in a region that contains 12 states within $\unit[5]{\icm}$, including degeneracies, whereas the third value is close to state 186 in a region where the density of states is significantly higher, namely 18 states within  $\unit[5]{\icm}$.
For the first two eigenstate computations we used a DMRG-based guess, whereas for 
the more difficult $\sigma=\unit[172.91]{\icm}$ we improved the initial guess by first computing a DMRG-based guess and then using it for seven inexact Lanczos iterations with $\sigma=\unit[179.22]{\icm}$.
After these seven Lanczos iterations, we used the resulting approximate TTNS eigenstate and its energy as a new $\sigma$ value. 
In addition to a fixed number of $5$ sweeps for the two low-energy computations,
the convergence parameters were set to those used for the $\bdimmax=10$  \ce{CH3CN} computations, 
see the SM Section S3.B for more information. 

\begin{figure}[!tbp]
\includegraphics[width=0.92\columnwidth]{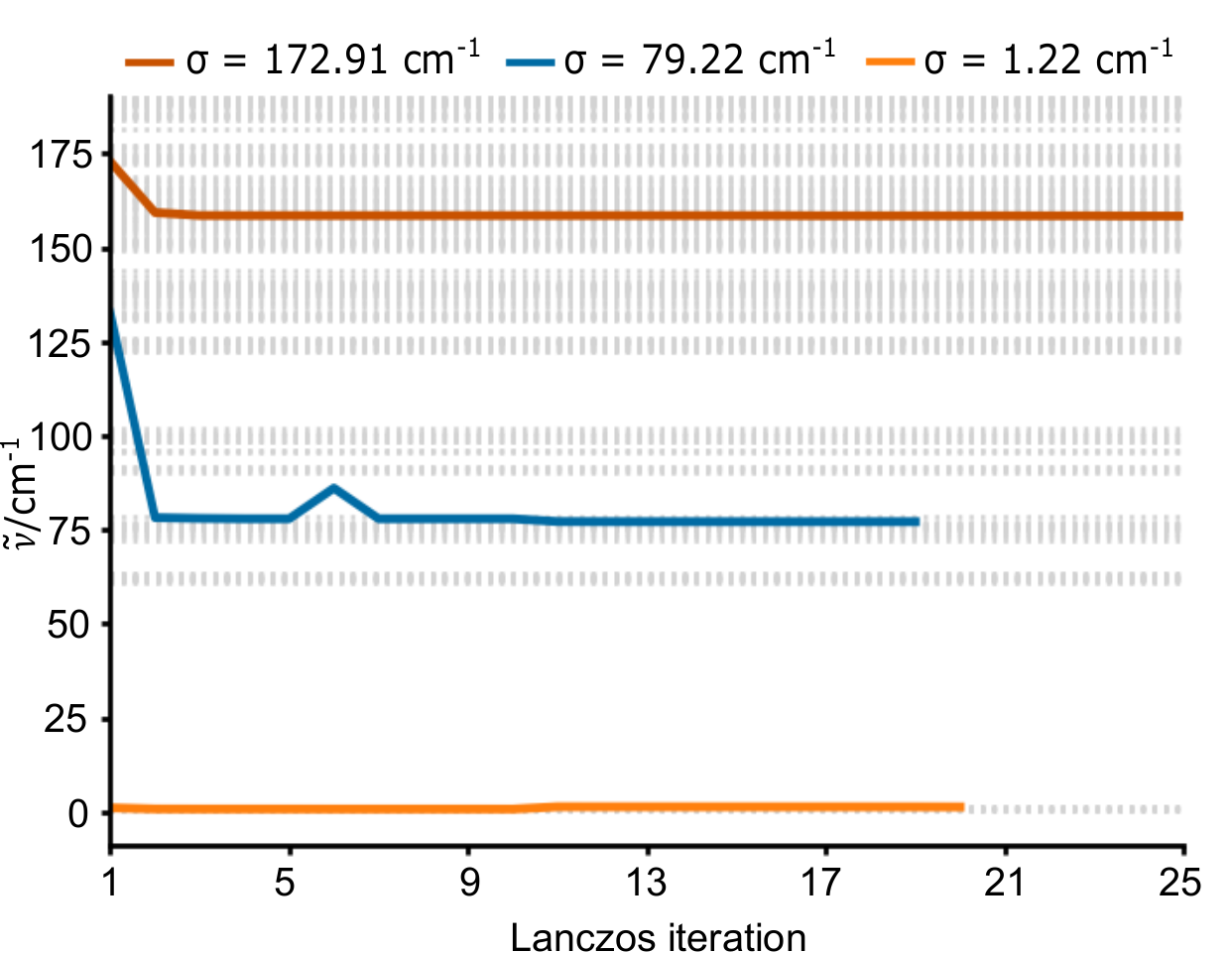}
\caption{TTNS energy convergences as a function of cumulative Lanczos iteration for the Eigen ion. Reference energies are shown as dotted gray lines.}
	\label{fig:eigen_conv}
\end{figure}

\begin{figure}[!tbp]
	\includegraphics[width=\columnwidth]{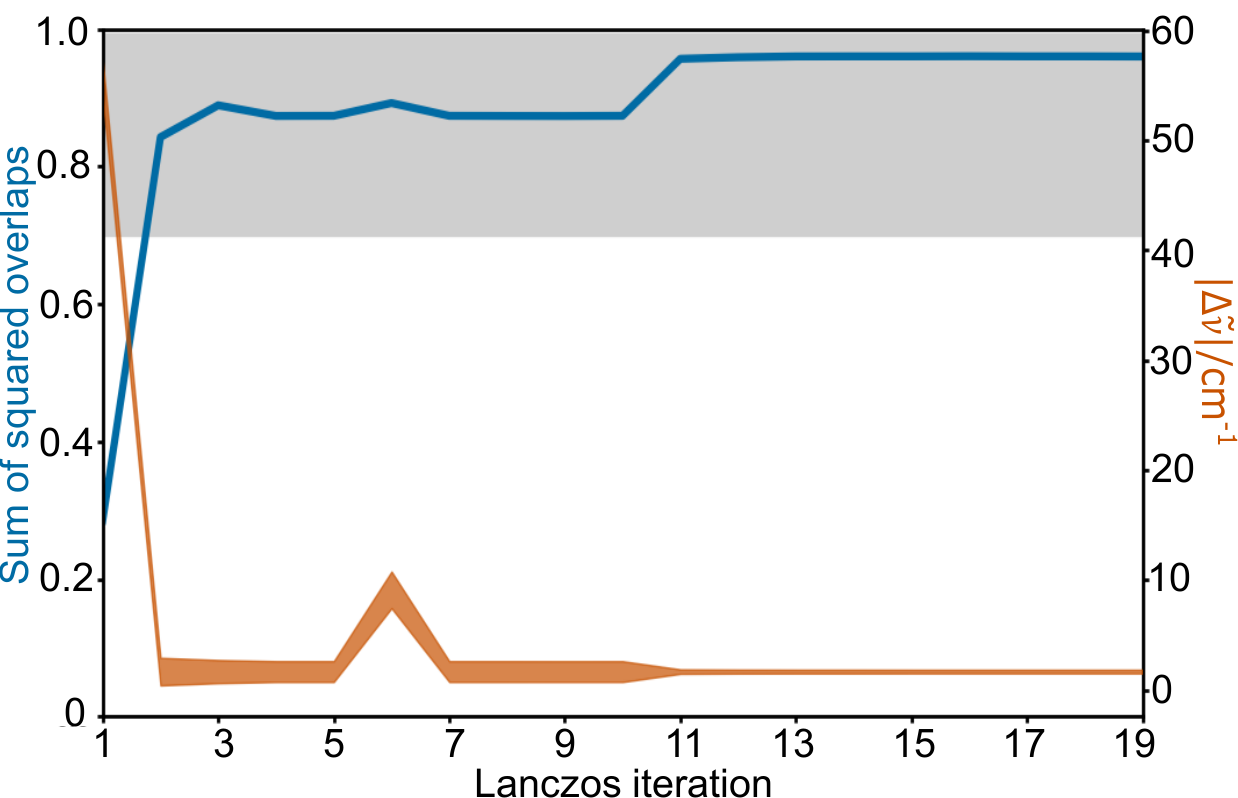}
	\caption{Convergence for the Eigen ion state with $\sigma=\unit[79.22]{\icm}$. The blue line and the left ordinate shows the sum of squared overlaps to DMRG-TTNS reference states in a small energy interval. The orange line and the right ordinate displays the absolute energy error, which is based on the used reference energy interval. 
	The gray region marks the convergence of the squared overlap to the approximate DMRG reference states. 
	See the text for details.}
	\label{fig:eigen_ovlp_err} %
\end{figure}

\begin{figure}[!tbp]
\includegraphics[width=\columnwidth]{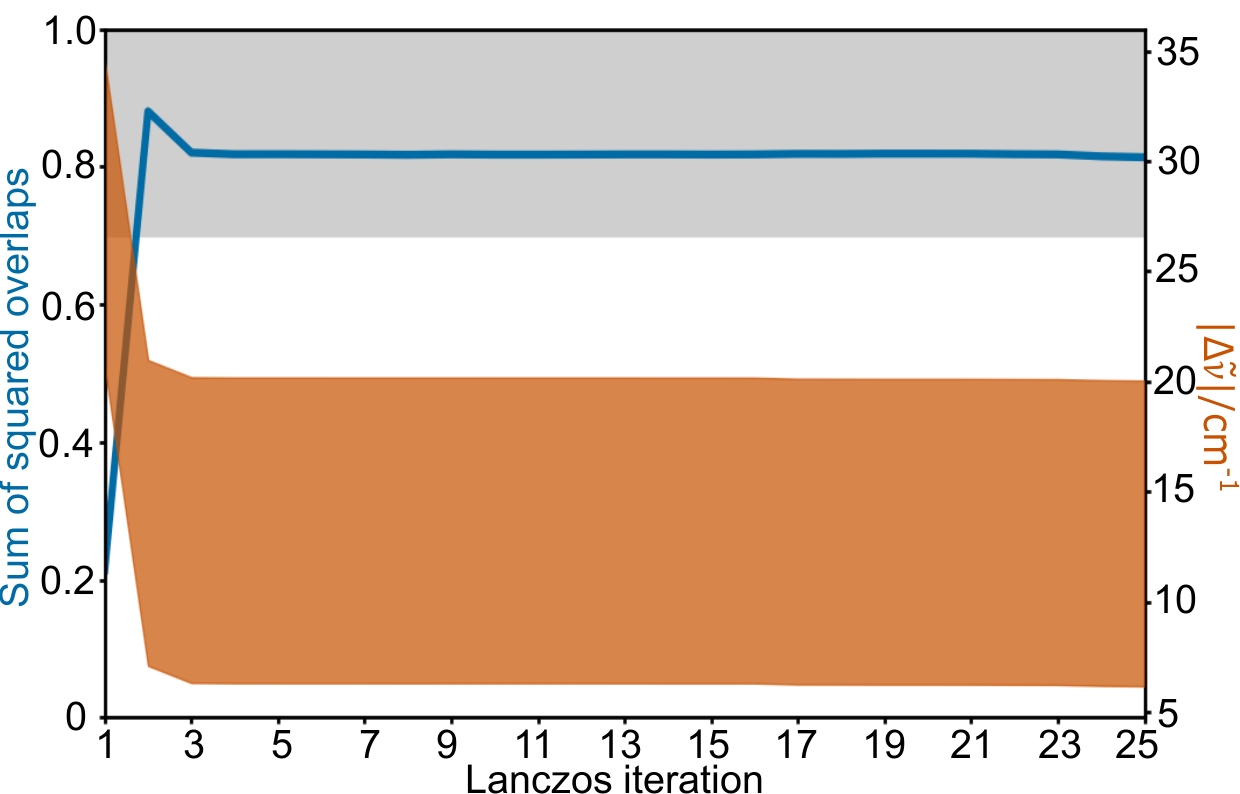}
\caption{Same as \autoref{fig:eigen_ovlp_err} but for $\sigma=\unit[172.91]{\icm}$.}
	\label{fig:eigen_ovlp_err_3206}
\end{figure}

The general convergence behavior is shown in \autoref{fig:eigen_conv}.
Importantly, due to the high density of states and complexity of the problem, the inexact Lanczos method does not necessarily find the state that is closest to $\sigma$ but rather a state that is close by. 
However, already after two  Lanczos iterations, the found states are converged to the eye.
The $\sigma=\unit[79.22]{\icm}$ computation displays some smaller convergence problems, which are resolved in the following iterations.
Overall, between $3$ and $11$ iterations are required to reach a relative convergence of $10^{-6}$.
To demonstrate stability over many  iterations,
more than $11$ iterations are shown in \autoref{fig:eigen_conv}
We now give a more detailed analysis of the convergence of the two high-energy states.
To restrict ourselves to analyzing one particular eigenstate only and to deal with root flipping,
we tracked one eigenstate over the Lanczos iterations by first following it based on an energy criterion and, after 5 (1) iterations for  $\sigma=\unit[79.22]{\icm}$ ($\unit[172.91]{\icm}$) by selecting it based on the overlap to the states from the previous iterations. 
Compared to \ce{CH3CN} and the Zundel ion, estimating the error in this case is very difficult, as the density of states is very high, and both the Lanczos and the DMRG reference TTNSs with $\bdimmax=70$ are not fully converged.
This leads to the computed states being linear combinations of reference states,
and even the $\bdimmax=150$ reference energies do not fully resolve degeneracies.
In other words, our reference states most likely consist of  linear combinations of the exact eigenstates.
To resolve the approximate nature of not only the computed eigenstates but also the reference states, we computed the summed squared overlap of the TTNS inexact Lanczos states to all TTNS DMRG states with $\bdimmax=70$ in a given energy interval, which we made as small as possible.
For $\sigma=\unit[79.22]{\icm}$
we used an interval of $[75.23,78.62]\unit{\icm}$ %
and for $\sigma=\unit[172.91]{\icm}$  we used an interval of $[138.33,152.19]\unit{\icm}$. %
For these intervals, the summed squared overlaps and energy errors are shown in \autoref{fig:eigen_ovlp_err} and \ref{fig:eigen_ovlp_err_3206}.
After just one iteration,  %
the squared overlaps for both computations %
are already larger than $0.7$,
which is also the maximum deviation of the TNS DMRG $\bdimmax=70$ states to the $\bdimmax=150$ after diagonalization.
The final summed squared overlap values are 0.96 (0.81) %
for  $\sigma=\unit[79.22]{\icm}$ ($\unit[172.91]{\icm}$).
Due to the $\bdimmax$ restriction, we cannot expect larger overlaps, so the final states can be considered fully converged.
Likewise, the energy error has some uncertainty, but for $\sigma=\unit[79.22]{\icm}$, the max.~possible difference to the $\bdimmax=70$ reference is only $\unit[1.8]{\icm}$ after  12 iterations.
For the more complicated $\sigma=\unit[172.91]{\icm}$ state, the energy error is larger, and we estimate it to be between $6$ and $\unit[20]{\icm}$. Given the small $\bdimmax=70$ for this strongly correlated system, this is a relatively small error.

\section{Conclusions}
\label{sec:conclusions}
Compared to ground state computations, computing arbitrary excited states in regions with high state densities using nonlinear wavefunction ansätze such as tree tensor network states (TTNSs) is very difficult. Here, we provided an extension of the inexact Lanczos method using shifted-and-inverted Hamiltonians to TTNSs. As vector operations like addition are approximate for TTNSs, this requires several modifications to the usual inexact Lanczos method, e.g., using approximate orthogonalization of the Krylov space.
Unlike DMRG methods, the current approach is not yet of a black-box character, and it requires some parameter tuning, particularly because the DMRG-based solution of the linear system of equations does not necessarily converge. Nevertheless, the inexact Lanczos method directly targets excited states, which DMRG methods are not designed for.  Multiple improvements of the developed  TTNS inexact Lanczos approach are possible, as well as extensions, e.g., to the rational Krylov method.\cite{Rational1984ruhe}

Already at the current state,
the inexact Lanczos TTNS approach enabled simulations for some of the currently most challenging vibrational problems in molecular quantum dynamics.
Particularly, we computed overall 122 states %
in different energy intervals of the difficult 12-dimensional acetonitrile benchmark system. We further showed how to quickly compute Fermi resonance states of the 15-dimensional fluxional Zundel ion, even when the initial guess does not capture the physics of the converged eigenstates. Finally, we demonstrated the computation of excited states for the 33-dimensional, fluxional, and very correlated Eigen ion in regions with high state densities.  
This opens the door to use the TTNS inexact Lanczos approach to accurately compute highly excited vibrational states such as OH stretch excitations, intermolecular excitations in molecular clusters, and vibronic excitations.
Importantly, our approach is not limited to vibrational systems. Given the generality of tensor network methods, we believe that the presented inexact Lanczos approach will also be useful to compute excited states of other many-body systems, e.g., in electronic structure theory.

\section*{Supplementary material}
See supplementary material for additional parameters, simulation details, and results.

\if\USEACHEMSO1
\begin{acknowledgement}
\else
\acknowledgements
\fi
This work was supported by the US
National Science Foundation (NSF) via grant no.~CHE-2312005.
This research was conducted using the Pinnacles cluster (NSF MRI, no.~2019144)
using CENVAL-ARC compute resources on the Pinnacles cluster (NSF no.~2346744) 
at the Cyberinfrastructure and Research Technologies (CIRT) at University of California, Merced.
\if\USEACHEMSO1
\end{acknowledgement}
\fi

\section*{Author Contributions}

\textbf{Madhumita Rano}
Data curation (equal); Formal analysis (equal); 
Investigation (equal); 
Software (equal);
Validation (equal);
Visualization (equal);
Writing -- original draft (supporting).
\textbf{Henrik R.~Larsson:} 
Conceptualization (lead);
Data curation (equal); 
Formal analysis (equal); 
Funding acquisition (lead); 
Investigation (equal); 
Methodology (lead);
Project Administration (lead);
Resources (lead);
Software (equal);
Supervision (lead); 
Validation (equal);
Visualization (equal);
Writing -- original draft (lead);
Writing – review \& editing (lead).

\section*{Data Availability}
The data that supports the findings of this study are available within the article. %

\if\USEACHEMSO1
\clearpage
\fi
\end{document}